%% file: ALLSMOG_paper.tex
\documentclass[useAMS,usenatbib]{mn2e}

\usepackage{amsmath,natbib,graphicx}
\usepackage{epsf}
\usepackage{epstopdf}
\input{epsf}
\usepackage{epsfig}
\usepackage{color}
\usepackage{subfigure}

\DeclareGraphicsExtensions{.jpg,.pdf,.png,.eps,.ps}

\newcommand{\HI}{H\:\!\textsc{i}}

\newcommand{\conv}{$\alpha_{\rm CO}$}

\def\simgt{\mathrel{\raise0.35ex\hbox{$\scriptstyle >$}\kern-0.6em
\lower0.40ex\hbox{{$\scriptstyle \sim$}}}}
\def\lta{\mathrel{\raise0.35ex\hbox{$\scriptstyle <$}\kern-0.6em
\lower0.40ex\hbox{{$\scriptstyle \sim$}}}}

\def\gs{\mathrel{\raise0.35ex\hbox{$\scriptstyle >$}\kern-0.6em
\lower0.40ex\hbox{{$\scriptstyle \sim$}}}}
\def\ls{\mathrel{\raise0.35ex\hbox{$\scriptstyle <$}\kern-0.6em
\lower0.40ex\hbox{{$\scriptstyle \sim$}}}}

\title[ALLSMOG Paper I]{ALLSMOG: an APEX Low-redshift Legacy Survey for MOlecular Gas \\ I -- molecular gas scaling relations, and the effect of the CO/H$_2$ conversion factor}
\author[M.\,S.\ Bothwell et al. ]
{M. S. Bothwell$^{1,2}$\thanks{E-mail:
matthew.bothwell@gmail.com},
J. Wagg$^{1,2,3,4}$, 
C. Cicone$^{1,2}$, 
R. Maiolino$^{1,2}$,
P. M\o ller$^{4}$, \newauthor
M. Aravena$^{4, 5}$,
C. De Breuck$^{6}$,
Y. Peng$^{1,2}$, 
D. Espada$^{7,8,9}$,
J. A. Hodge$^{10}$, \newauthor
C. M. V. Impellizzeri$^{7,10}$, 
S. Mart\'in$^{11}$,
D. Riechers$^{12}$, 
F. Walter$^{13}$ \\
$^{1}$Cavendish Laboratory, University of Cambridge, 19 J.J. Thomson Avenue, Cambridge, CB3 0HE, UK\\
$^{2}$Kavli Institute for Cosmology, University of Cambridge, Madingley Road, Cambridge CB3 0HA, UK\\
$^{3}$Square Kilometre Array Organisation, Jodrell Bank Observatory, Lower Withington, Macclesfield, Cheshire SK11 9DL, UK\\
$^{4}$European Southern Observatory, Alonso de C\'ordova 3107, Vitacura, Casilla 19001, Santiago 19, Chile\\
$^{5}$N\'ucleo de Astronom\'{\i}a, Facultad de Ingenier\'{\i}a, Universidad Diego Portales, Av. Ej\'ercito 441, Santiago, Chile\\
$^{6}$European Southern Observatory, Karl Schwarzschild Stra§e 2, 85748 Garching bei M\"{u}nchen, Germany\\
$^{7}$Joint ALMA Observatory, Alonso de Cordova 3107, Vitacura, 763-0355, Santiago, Chile\\
$^{8}$National Astronomical Observatory of Japan (NAOJ), 2-21-1 Osawa, Mitaka, 181-8588, Tokyo, Japan\\
$^{9}$The Graduate University for Advanced Studies (SOKENDAI), 2-21-1 Osawa, Mitaka, 181-8588, Tokyo, Japan\\
$^{10}$National Radio Astronomy Observatory, 520 Edgemont Rd, Charlottesville, VA, 22903, USA\\
$^{11}$Institut de Radio Astronomie Millim\'etrique, 300 rue de la Piscine, Dom. Univ., 38406, St. Martin d'H\`eres, France\\
$^{12}$Department of Astronomy, 220 Space Science Building, Cornell University, Ithaca, NY, 14853, USA\\
$^{13}$Max Planck Institut f\"{u}r Astronomie, K\"{o}nigstuhl 17, 69117 Heidelberg, Germany\\
}
\begin{document}
\date{Accepted ----. Received ---- in original form ----}

\pagerange{\pageref{firstpage}--\pageref{lastpage}} \pubyear{2014}

\maketitle
\begin{abstract}
We present ALLSMOG, the  APEX Low-redshift Legacy Survey for MOlecular Gas. ALLSMOG is a survey designed to observe the CO($2-1$) emission line with the APEX telescope, in a sample of local galaxies ($0.01 < z < 0.03$), with stellar masses in the range 8.5 $<$ log(M$*$/M$_{\sun}$) $<$ 10. This paper is a data release and initial analysis of the first two semesters of observations, consisting of 42 galaxies observed in CO($2-1$). 

By combining these new CO$(2-1)$ emission line data with archival \HI\ data and SDSS optical spectroscopy, we compile a sample of low-mass galaxies with well defined molecular gas masses, atomic gas masses, and gas-phase metallicities. We explore scaling relations of gas fraction and gas consumption timescale, and test the extent to which our findings are dependent on a varying CO/H$_2$ conversion factor. We find an increase in the H$_2$/\HI\ mass ratio with stellar mass which closely matches semi-analytic predictions. We find a mean molecular gas fraction for ALLSMOG galaxies of M$_{\rm H2}$/M$_*$ =  ($0.09 - 0.13$), which decreases with stellar mass. This decrease in total gas fraction with stellar mass is in excess of some model predictions at low stellar masses. We measure a mean molecular gas consumption timescale for ALLSMOG galaxies of $0.4 - 0.7$ Gyr. We also confirm the non-universality of the molecular gas consumption timescale, which varies (with stellar mass) from $\sim$100 Myr to $\sim$2 Gyr. Importantly, we find that the trends in the H$_2$/\HI\ mass ratio, gas fraction, and the non-universal molecular gas consumption timescale are all robust to a range of recent metallicity-dependent CO/H$_2$ conversion factors. \\ \\

\end{abstract}

\begin{keywords}
galaxies: evolution -- 
galaxies: formation -- 
galaxies: abundances --
galaxies: statistics
\end{keywords}

\section{Introduction}
\label{sec:introduction}

The cold phase of the interstellar medium (ISM) is a central ingredient in the galaxy evolution process. In the early Universe, rarified cold gas rained down into the potential wells of primordial dark matter haloes, but it was not until the gas collapsed and fragmented, forming dense molecular clouds, that star formation could begin. In the present-day Universe stars form exclusively in the dense cores of molecular clouds, and, as such, molecular gas acts as the critical precursor for star formation.   

In spite of the importance of molecular gas in driving star formation, there remain many outstanding problems to be addressed by extragalactic surveys. While modern optical galaxy surveys boast sample sizes into the millions, and the latest surveys for extragalactic {\it atomic} gas (in the form of \HI) have many tens of thousands of members, typical surveys for molecular gas generally detect, at most, on the order of a few hundred galaxies. A primary difficulty in observing molecular hydrogen is its lack of a permanent dipole moment, making H$_2$ exceptionally faint in emission, directly detectable in only a handful of very local sources. Combined with the fact that the IR emission lines of H$_2$ are emitted by warm gas (rather than tracing the bulk of the cold molecular ISM), this makes a direct detection of H$_2$ an utterly impractical strategy for a large extragalactic survey. 

The traditional approach to overcoming this issue is to use a `tracer' molecule, which is easily observable and can be assumed to be approximately co-spatial with cold H$_2$. By far, the most popular choice of tracer molecule in use is carbon monoxide ($^{12}$C$^{16}$O --  simply `CO' hereafter)\footnote{Using dust -- observable via the far-IR continuum -- as a gas tracer is also becoming popular; see \cite{2014A&A...562A..30S}; \cite{2014ApJ...783...84S}.}. CO makes for an ideal tracer molecule -- it has a low excitation energy, and exhibits bright emission lines at approximately integer multiples of 115.27 GHz, the two lowest transitions of which fall conveniently into the mm-wavelength atmospheric windows. CO is therefore detectable across an enormous distance range, from the very local Universe out to $z > 5$ (\citealt{2013ApJ...767...88W}; for a recent review see \citealt{2013ARA&A..51..105C}).  

The inherent difficulty in using CO to trace molecular gas, of course, is that {\it by mass} it represents only a tiny fraction of the molecular ISM (which is predominantly composed of molecular hydrogen and helium). Converting from a CO luminosity to a total mass of molecular gas, therefore, requires the use of a `conversion factor' (often denoted `\conv '), which encodes information about the fraction of CO relative to H$_2$. There has been a vast amount of work over the last few decades dedicated to empirically measuring, and theoretically modelling, \conv\ (see \citealt{2013ARA&A..51..207B} for a recent review) for individual molecular clouds. By making a few assumptions (such as a population of isolated and virialised molecular clouds), it is possible to derive a {\it global} \conv, applicable -- in an average sense -- to entire galaxies. 

This assumed value of \conv\ is not a universal constant however -- the value for any specific galaxy may be higher or lower than the average value estimated for the Milky Way, depending on the specific conditions within the ISM. It is now generally accepted that the primary factor driving galaxy-to-galaxy variations in \conv\ is the metal abundance of the ISM. A high abundance of metals in a molecular cloud results in more dust, which can shield CO from photodissociation. At lower metallicities, the region of a molecular cloud capable of forming CO molecules shrinks further and further into the dense core of the cloud. As such, a molecular gas cloud will produce less and less CO emission (for a given H$_2$ mass) as metallicity is decreased. This holds true when considering the integrated properties of galaxies -- galaxies with lower gas-phase metallicities require higher values of \conv\ in order to accurately derive their molecular gas masses (e.g., \citealt{Genzel2011aa}; \citealt{2011AJ....142...37S}; \citealt{2012MNRAS.421.3127N}). 

This metallicity-dependence of \conv\ may have limited extragalactic CO surveys in two ways.  Firstly, due to the strong link between metallicity and stellar mass (e.g. \citealt{2004ApJ...613..898T}), molecular gas becomes increasingly difficult to detect in low mass galaxies. As such, most large surveys for molecular gas have been preferentially targeted towards samples at high stellar masses, while the molecular gas content of galaxies with low stellar masses, $M_* \ls 10^{10} M_{\sun}$ -- a group which contains much of the molecular gas in the present day Universe \citep{2003ApJ...582..659K} -- remains comparatively poorly explored. The second difficulty is that without accurate gas-phase metallicities, molecular gas observations are difficult to interpret, necessitating the unphysical assumption of a constant \conv\  which can lead to underlying trends being obscured or biased.

There have been a number of surveys undertaken to detect molecular gas, both locally and in the high-redshift Universe. Locally, observations tend to focus on lower-$J$ transitions of CO in samples of massive `main sequence' galaxies (\citealt{2011MNRAS.415...32S}; \citealt{2014arXiv1401.7773B}; though see \citealt{2005ApJ...625..763L} for a smaller sample of low-mass galaxies), whereas gas observations at high redshift have, until recently, concentrated on higher-$J$ observations of bright, starburst galaxies (e.g., \citealt{2005MNRAS.359.1165G}; \citealt{2006ApJ...640..228T}; \citealt{Bothwell:2010aa}; \citealt{2013MNRAS.429.3047B}; for a recent review see \citealt{2013ARA&A..51..105C} and references therein)\footnote{Technological advances have improved both of these shortcomings, however, with CO observations of high-$z$ `main sequence' galaxies being undertaken \citep{2013ApJ...768...74T}, as well as low-$J$ observations at high-$z$ (i.e., \citealt{Ivison:2011aa}; \citealt{2011ApJ...733L..11R}) enabled by the addition of (26.5 to 40 GHz) Ka-band receivers on sensitive telescopes like the Karl G. Jansky Very Large Array (VLA).}. 

In recent years, there has been an effort to substantially increase the statistics of low-redshift molecular gas observations by means of large dedicated surveys: two such surveys are COLDGASS (the CO Legacy Database for the Galaxy Evolution Explorer (GALEX) Arecibo Sloan Digital Sky Survey (SDSS) Survey), and the Herschel Reference Survey. COLDGASS  \citep{2011MNRAS.415...32S} is the molecular gas extension of the \HI-based Galex-Arecibo SDSS survey `GASS' \citep{2010MNRAS.403..683C}. COLDGASS is designed to survey molecular gas in nearby massive (M$* > 10^{10}$ M$_{\sun}$) galaxies, selected to have both SDSS and GALEX UV coverage. COLDGASS has successfully explored molecular gas scaling relations and gas consumption timescales at high stellar masses, finding (amongst other results) strong connections between the gas content and galaxy colour/morphology, and a non-universal molecular gas depletion timescale \citep{2011MNRAS.415...61S}. The Herschel Reference Survey (HRS hereafter; \citealt{2010PASP..122..261B}) is one program designed to start overcoming the bias towards high stellar masses. \cite{2014arXiv1401.7773B} obtained new molecular gas observations for 59 galaxies (and archival CO data for a further 166), with stellar masses as low as $10^9 \;{\rm M}_{\sun}$, and explored a variety of scaling relations in this low stellar mass regime \citep{2014arXiv1401.8101B}. Using a H-band-dependent \conv,  \cite{2014arXiv1401.8101B} find that several scaling relations (including the molecular gas consumption timescale and the H$_2$/HI ratio) become statistically insignificant, or even disappear completely, when using this varying \conv.

Mindful of the bias in molecular gas surveys towards samples of massive galaxies, we designed ALLSMOG, the Apex Large Legacy survey for MOlecular Gas. The ALLSMOG survey aims to explore molecular gas scaling relations, gas fractions, atomic/molecular gas ratios, and gas consumption timescales in a sample of `normal' low mass local galaxies, with well-defined metallicities and stellar masses in the range 8.5 $<$ log(M$*$/M$_{\sun}$) $<$ 10. Requiring a well-defined metallicity for inclusion in the ALLSMOG sample will ensure that it is possible to explore the effect of a variety of metallicity-dependent CO/H$_2$ conversion factors on any derived scaling relations, and -- critically -- identify the correlations that persist independent of the choice of conversion factor.

ALLSMOG is an APEX large program, being awarded 300 hours over 4 semesters, and is due to be completed in 2015. To date, the ALLSMOG program is $\sim 40\%$ complete after two semesters of observing, and has obtained CO($2-1$) observations for a total of 42 galaxies. This paper presents a data release and analysis of this first $\sim 40\%$ of our sample. The ALLSMOG survey was intended as a legacy survey for molecular gas in low-mass galaxies -- as such, all reduced spectra have been made publicly available to the community, and can be accessed at {\it www.mrao.cam.ac.uk/ALLSMOG}.

Throughout this work, we have assumed a standard flat $\Lambda$-CDM cosmology with H$_0$ = 70 km$^{-1}$s$^{-1}$Mpc.

\section{Sample selection}
\label{sec:sample}

ALLSMOG is designed to survey molecular gas in local galaxies with stellar masses M$* < 10^{10}$ M$_{\sun}$ -- a mass range that is poorly explored by existing molecular gas surveys.  We opted to survey the CO($2-1$) emission line, as it traces the cold molecular reservoir essentially as effectively as the CO($1-0$) line, while benefiting from a far higher expected flux density.

Our survey takes advantage of the enormous wealth of photometric and spectroscopic data made public as part of the SDSS data release 7. The basic SDSS DR7 data have been analysed by groups at the Max-Planck-Institut f\"{u}r Astrophysik (MPA) and JHU to produce catalogues\footnote{http://www.mpa-garching.mpg.de/SDSS/DR7/} with derived physical properties (stellar masses, star-formation rates, metallicities, etc.), which are briefly described in \S \ref{sec:ancilliary} below. The SDSS stellar masses are derived based on fitting described in \cite{2003MNRAS.341...54K}, and star formation rates are based on methods detailed in \cite{2004MNRAS.351.1151B}.

To assemble the ALLSMOG survey catalogue, we make cuts to the parent sample of the MPA-JHU catalogue to include galaxies:
\begin{enumerate}
\item{with stellar masses, $8.5 < \log(M_*/M_{\sun}) < 10.0$;}
\item{with declinations less than $+10^\circ$ (to ensure that they are observable with APEX at high elevations);}
\item{with accurately measured metallicities (see \S 3.2) 12+log(O/H) $> 8.5$ (where the CO-to-H2 conversion is expected to be similar to, or lower than the Milky Way Value);}
\item{lying in the redshift range, $0.01 < z < 0.03$}
\item{with an existing \HI\ observation from the literature.}
\end{enumerate}

After applying these cuts to the parent SDSS spectroscopic catalogue, the final target list was determined based on the targets' position on the sky, taking into account both the position of the Sun (targets too close to the Sun cannot be observed) and the position of potential flux calibrators. 

The lower redshift bound of the final selection criterion has been chosen so as to ensure that the $\sim 27''$ APEX beam at 230 GHz covers the area of CO emission. Galaxies observed at lower redshifts have a high probability that their CO line emission will be larger than the size of the beam. As a final check, we manually checked the optical extent of all galaxies to ensure that they fall approximately within the APEX beam (the few galaxies which have slightly larger optical extents are still small enough to have most of their CO emitting regions within the beam -- see \S4.0.1 for discussion). The high-redshift cutoff has been chosen so as to keep the integration times reasonable given the expected faintness of low-J CO line emission in more distant objects (particularly considering the low stellar masses and metallicities of our sample).

We note that our selection criteria on the metallicity is sufficient to cover the expected range in metallicities for SDSS galaxies over our stellar mass range, $8.5 < \log(M_*) < 10.0$. As discussed above, lower metallicity galaxies typically have increased values of $\alpha_{\rm CO}$ and therefore the detection of CO($2-1$) line emission becomes unfeasible for APEX. Our selection criteria leaves us with a sample of late-type galaxies which have a range of estimated star-formation rates $(0.01 < \log ({\rm SFR/M}_{\sun}\, {\rm yr}^{-1}) < 2.5)$, and metallicities $(8.5 < [12 + \log {\rm O/H}] < 9.2)$. %In addition, the optical emission line properties of our galaxies classify them as both starburst galaxies and AGN, so we can explore any differences that may exist in the gas mass fractions of these two populations. 

Table B1 lists the physical properties of the galaxies in the ALLSMOG sample. 
 
\section{Ancillary observations}
\label{sec:ancilliary}

\subsection{HI overlap}

At the relatively low stellar masses probed by our survey, it is likely that galaxies will have substantial (or even dominant) contributions to their total gas content coming from \HI\ (e.g. \citealt{2009MNRAS.400..154B}; \citealt{2011MNRAS.418.1649L}). In order to be able to measure the {\it total} gas content (H$_2$+\HI), as well as explore any trends in the H$_2$/\HI\ ratio, we ensured that every member of the ALLSMOG sample had an archival \HI\ observation, generally from either the HI Parkes All Sky Survey (HIPASS) \citep{2004MNRAS.350.1195M}, The Arecibo Legacy Fast ALFA Survey (ALFALFA) \citep{2011AJ....142..170H}, or -- failing either of these -- from the large collection of \HI\ observations assembled by \cite{2005ApJS..160..149S}. Given our sample selection was from the SDSS spectroscopic survey, the majority of the ALLSMOG sample also appears in the ALFALFA survey. 

\subsection{Deriving physical parameters}

%%%%%%%% FIG 1 %%%%%%%
\begin{figure*}
\centering
\mbox
{
\includegraphics[width=17cm]{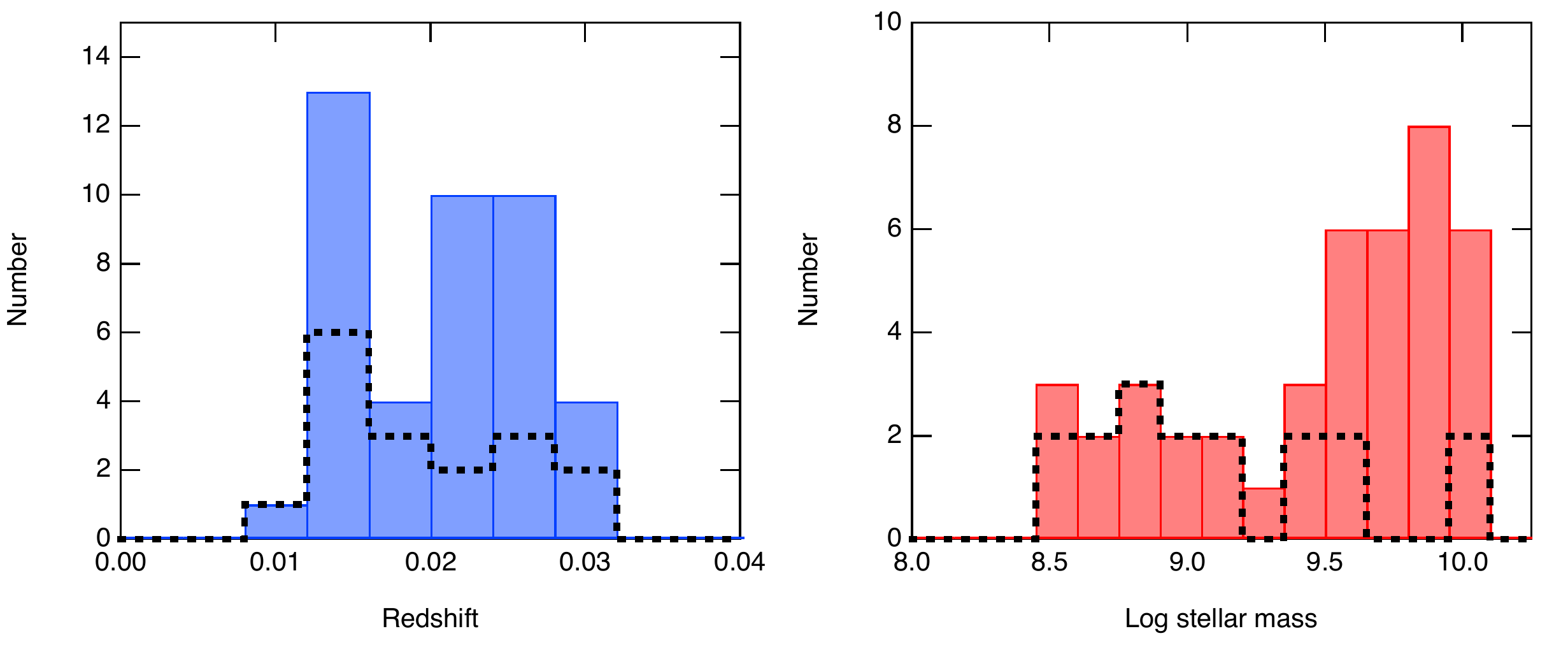}	
}
\caption{Histograms of the numbers of galaxies in the ALLSMOG sample, showing the distribution in distance ({\it Left Panel}) and stellar mass ({\it Right Panel}). In each, the full ALLSMOG sample is traced by a coloured histogram, while non-detections are traced by a dashed line. While the distribution in redshift is relatively independent of detection status, galaxies at lower stellar masses have a poorer detection rate than those at higher stellar masses.}
\label{fig:hist}
\end{figure*}
%%%%%%%%%%%%%%%%%%%

Our sample galaxies are selected from the SDSS spectroscopic survey, and have their stellar masses and star formation rates readily available. We adopt both stellar masses and star formation rates (corrected for fibre aperture effects) as listed in the MPA-JHU catalogue. 

We derive gas-phase metallicities using the available optical strong-line fluxes, and the calibration given in \cite{2008A&A...488..463M}, which uses a combination of models and empirical calibration. Specifically, we use two diagnostics, the N[{\sc ii}]/H$\alpha$ ratio and the `R23' parameter (= $(\rm [OII]3727 \, + \, \rm [OIII]4958,5007) / H\beta$), and derive a metallicity based on each. We take as our final metallicity measurement the mean of the two derivations: any galaxies for which the two methods produce metallicities which are discrepant by $>0.2$ dex were discarded at the sample-selection phase, and do not appear in our final sample.

\section{Observations and data reduction}

Observations of the CO($2-1$) line were carried out during 2013 and 2014 using the 230 GHz Swedish Heterodyne Facility Instrument (SHFI; \citealt{2008A&A...490.1157V}) on the APEX telescope, located at Llano Chajnantor in Chile. Observing conditions over the two semesters varied from relatively poor to excellent (precipitable water vapour $0.2 < ({\rm PWV/mm}) < 3$), with typical average conditions of PWV$\sim 2$mm. Focusing was determined using planets (Jupiter or Saturn) approximately every 2 hours. Pointing corrections were carried out once per hour using a bright quasar close to the science target.%, leading to a typical astrometric uncertainty of XX$''$.  

Data were reduced using the {\sc class} package in {\sc Gildas}\footnote{http://www.iram.fr/IRAMFR/GILDAS/}. All observations were examined by hand, scan by scan, and any individual scans displaying significant distortions, like baseline ripples, were discarded. Typical final spectral noise levels (after removing bad scans) were $\sim 2$mK per 20 km\,s$^{-1}$ channel. For consistency and ease of use, we hereafter convert all flux and noise values from Kelvin to Janskys: the APEX-1 receiver has a telescope efficiency at 230 GHz of 39 Jy/K, estimated using the Ruze formula. Typical spectral noise levels therefore, in mJy, are $\sim 80$ mJy per 20 km\,s$^{-1}$ channel. (Noise levels for individual observations are listed in Table \ref{tab:obs}.) We also calculate an aperture correction factor, to correct for potential CO flux falling outside the $27''$ beam -- in the majority of cases, essentially all the CO flux falls within the beam (the mean amount of flux falling inside the beam is $92 \pm 7\%$). Appendix Section \ref{sec:aper} gives more details of this correction factor. Individual correction factors are listed in Tab. B2, and all CO luminosities and H$_2$ masses hereafter are corrected for aperture effects. 

Final reduced spectra were binned to 20 km\,s$^{-1}$ resolution, and further analysed in IDL. CO($2-1$) flux densities were measured by velocity-integrating Gaussian profiles fit to the central emission line. Gaussian profiles were chosen in order to simultaneously measure the line width, flux density, and any potential continuum contribution. All but two sources were well-ft by single Gaussian profiles: the two exceptions, 2MASJ0846+02 and NGC2936, had significant non-Gaussian emission and were poorly fit by a single profile. For these sources, we fit double Gaussian profiles (Figs B1 and B3). CO luminosities were then calculated using the standard relation given by \cite{Solomon:2005aa}:

\begin{equation}
 L'_{\mathrm {CO}} = 3.25 \times 10^7 \; S_{\mathrm {CO}} \Delta v  \; \nu_{\mathrm {obs}}^{-2} \; D_L^2 \; (1+z)^{-3},
\end{equation}

where $L'_{\mathrm {CO}}$ is the line luminosity in K km\,s$^{-1}$ pc$^{2}$, $S \Delta v$ is the velocity-integrated CO line flux in Jy km\,s$^{-1}$ , $\nu_{obs}$ is the observed frequency of the line in GHz, and $D_L$ is the distance in Mpc. For the purposes of this work, we assume that the CO($2-1$) line is fully thermalised -- that is, we assume $T_b(2-1)/T_b(1-0)=1$ (where $T_b$ is the equivalent Rayleigh-Jeans brightness temperature in excess of that of the cosmic microwave background). Making the assumption that the CO($2-1$) line is slightly sub-thermalised would raise our final derived CO($1-0$) line luminosities by approximately 10\% (for a typical brightness temperature ratio). 

For sources not detected in CO, we calculate 3-sigma upper limits on the CO flux density based on the measured rms channel noise:

\begin{equation}
\mathrm{S}_{CO} < 3 \sigma \; \sqrt{\Delta V_{CO}\;  dv}\;, 
\end{equation}
where $\sigma$(Jy) is the channel noise given in Table B2, $\Delta V_{CO}$ is the mean linewidth of the detected sample (= 130 km\,s$^{-1}$), and $dv$ is the bin size in km\,s$^{-1}$ (= 20 km\,s$^{-1}$). 

Of the 42 sources observed to date, we detect 25 in CO($2-1$) emission -- a detection rate of  60\%. Figure \ref{fig:hist} shows histograms of redshift and stellar mass for the ALLSMOG sample, with CO detections and non-detections separated. It can be seen that there is no apparent redshift bias in our detection rates (Figure \ref{fig:hist}, left panel): a Kolmogorov-Smirnov test of the redshifts of the detected and non-detected sub-samples returns a P value of 0.59, suggesting that there is no statistical difference between the two. 

There is, on the other hand, a strong {\it stellar mass} bias in the detection rate (Fig. \ref{fig:hist}, right panel). Galaxies at lower stellar masses become increasingly difficult to detect in CO emission, with the result that essentially all galaxies below $10^9$ M$_{\sun}$ are non-detections. The exception, NGC2936 (which has a stellar mass log(M$_*$) = 8.5 M$_{\sun}$), is undergoing a major merger (see appendix Fig. A3), and is significantly elevated above the `main sequence' of star formation. Full details of our observations are listed in Table \ref{tab:obs}.

\subsection{Deriving H2 masses}
\label{sec:h2}

Molecular hydrogen masses can be calculated from the CO luminosity, $L'_{\rm CO}$, by assuming a CO/H$_2$ conversion factor (which we refer to as $\alpha_{\rm CO}$):

\begin{equation}
{\rm M(H}_2) = \alpha_{\rm CO}L'_{\rm CO}
\end{equation}

As discussed above, while a number of physical factors can cause $\alpha_{\rm CO}$ to vary, it is likely that one of the dominant factors is metallicity, with $\alpha_{\rm CO}$ increasing as metallicity decreases. As such, throughout this work we will present results assuming two cases:

\begin{enumerate}
\item{A constant, Milky-Way appropriate conversion factor of $\alpha_{\rm CO}$ = 4.5 M$_{\sun} \;({\rm K\; km\,s^{-1} \;pc}^2)^{-1}$; \citep{2013ARA&A..51..207B}. The units of $\alpha_{\rm CO}$ are hereafter omitted. This value includes a correction of 1.36 to account for interstellar helium.}
\item{A conversion factor which is derived galaxy by galaxy, dependent on the gas-phase metallicity. }
\end{enumerate}

There are a number of metallicity-dependent CO/H$_2$ conversion factors in the literature (\citealt{2013ARA&A..51..207B} presents a compilation of several recent ones). In the interest of brevity, we will not present all plots for every available conversion factor. Instead, we will plot results assuming both a constant \conv, and the metallicity-dependent conversion factor derived by \cite{2010ApJ...716.1191W}.% \cite{2010ApJ...716.1191W} model the increase in \conv\ with decreasing metallicity as:
%
%\begin{equation}
%\log_{10} \alpha_{\rm CO} = a_1 \log_{10}(Z/Z_{\sun}) + a_2,
%\frac{\alpha_{\rm CO}(Z')}{\alpha_{\rm CO}(Z' = 1)} = \exp \frac{+4.0 \Delta A_V}{Z' < A_{V, MW} >} \exp \frac{-4.0 \Delta A_V}{< A_{V, MW} >},
%\end{equation}
%
%where $Z'$ is the abundance of heavy elements and dust (relative to solar), $\alpha_{\rm CO}(Z' = 1)$ is the CO/H$_2$ conversion factor at solar metallicity, $\Delta A_{V}$ is the difference in optical depth between H$_2$ and CO in a GMC,  and $< A_{V, MW} >$ is the mean extinction through a GMC at the metallicity of the MW. 
The  \cite{2010ApJ...716.1191W} conversion factor models \conv\ as relatively flat at high ($\sim$ solar) metallicities, with a sharply non-linear increase towards higher values of \conv\ as metallicity decreases. \cite{2013ARA&A..51..207B} note that the \cite{2010ApJ...716.1191W} \conv\ prediction provides the best fit to existing data \citep{2013ApJ...777....5S}. 

%%%%%%%% FIG 2 %%%%%%%
\begin{figure}
\centering
\mbox
{
  \includegraphics[width=8.5cm]{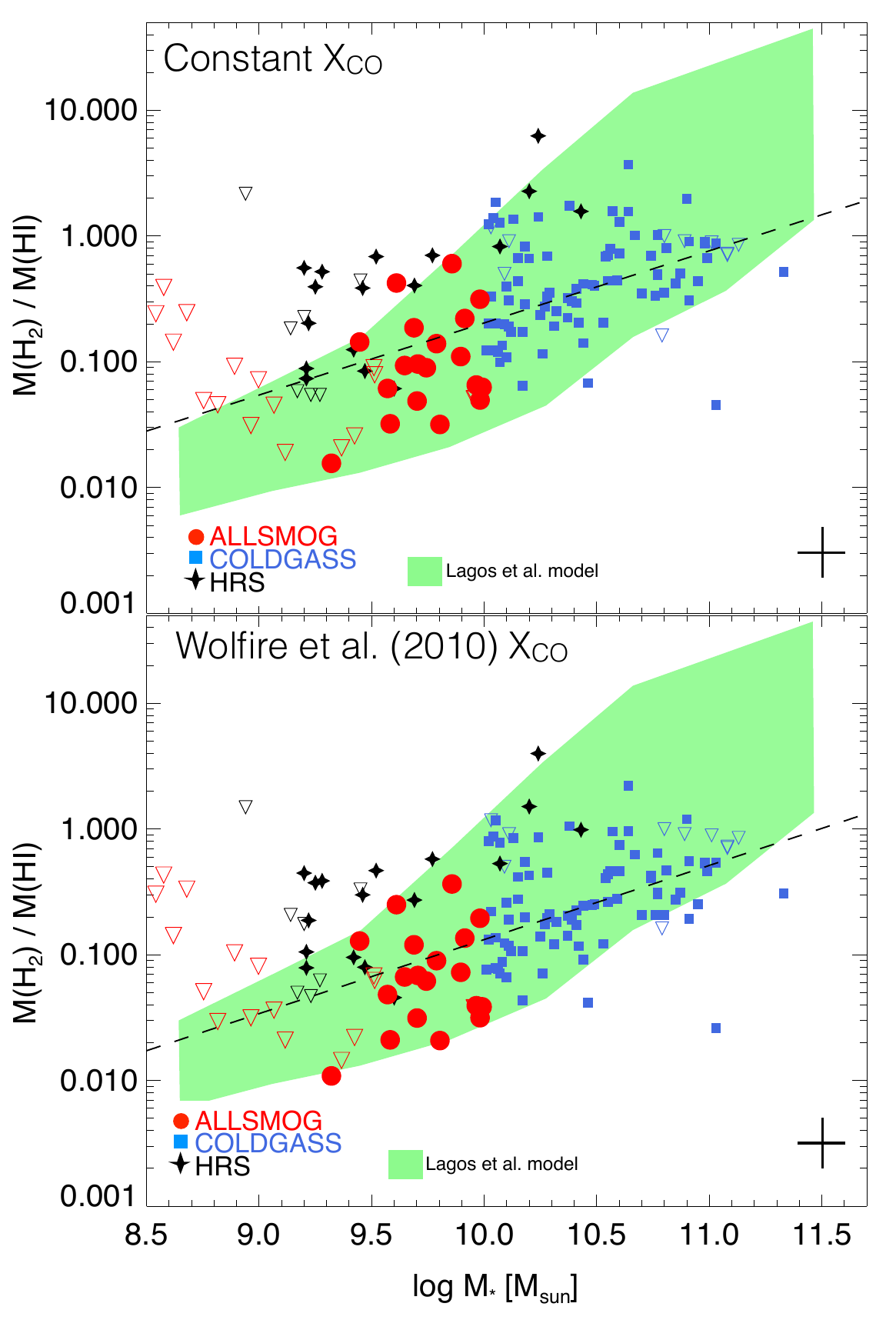}
}
\caption{H$_2$/\HI\ ratios, for the ALLSMOG sample. Also plotted are the galaxies from the COLDGASS and HRS surveys with well-defined metallicities (as described in the text). The panel has molecular gas masses calculated using a constant, Milky-Way $\alpha_{\rm CO}$. The lower panel has molecular gas masses calculated using the metallicity-dependent $\alpha_{\rm CO}$ of Wolfire et al. (2010). Filled symbols and open triangles denote detected sources and 3$\sigma$ upper limits, respectively. Black dotted lines show linear regression fits, estimated using a Bayesian MCMC technique to account for the upper limits. The green shaded region shows the semi-analytic model prediction from Lagos et al. (2011) (the upper and lower bounds of the coloured region mark the 90th and 10th percentiles of the model distribution).}
\label{fig:h2hi}
\end{figure}
%%%%%%%%%%%%%%%%%%%%

%%%%%%%% FIG 2b %%%%%%%
\begin{figure}
\centering
\mbox
{
  \includegraphics[width=8.5cm]{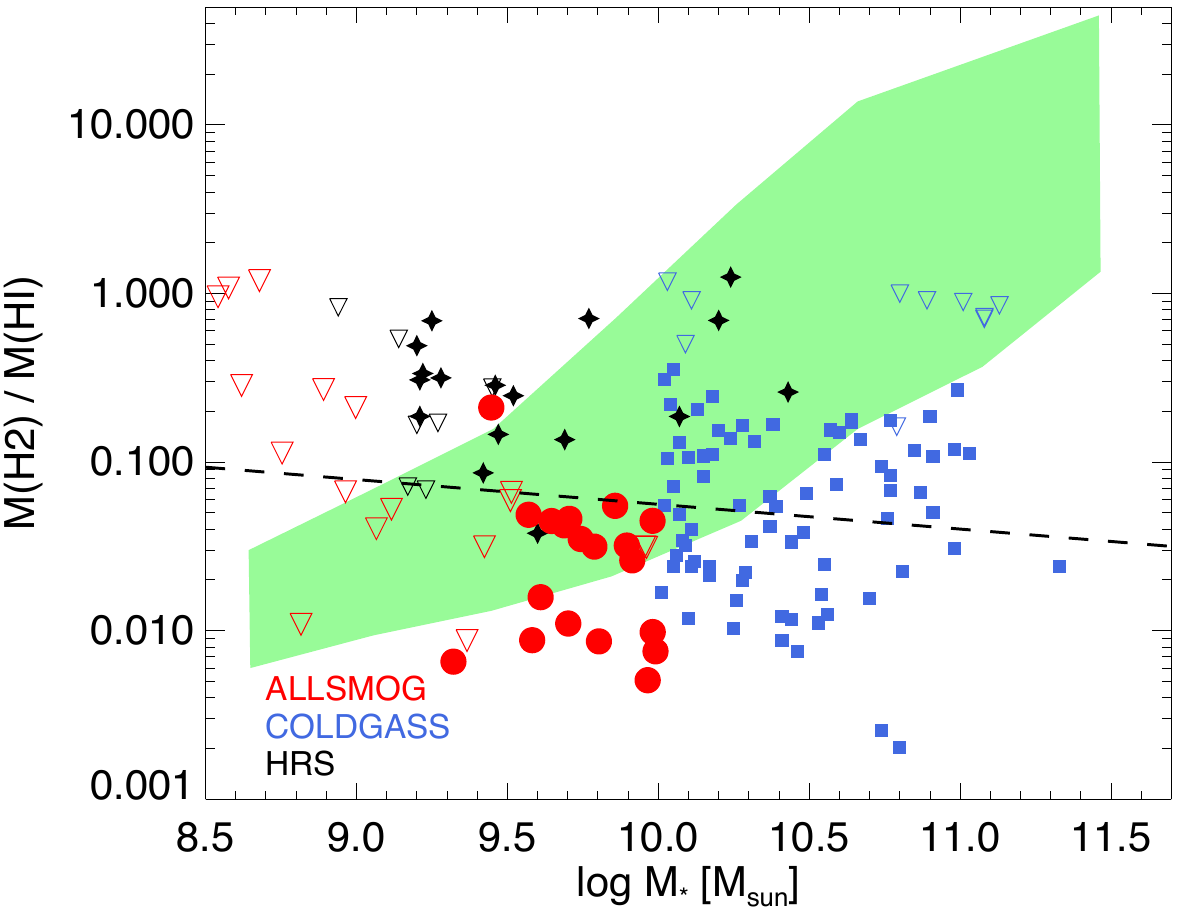}
}
\caption{H$_2$/\HI\ ratios, for the ALLSMOG sample, calculated using the Israel (1997) metallicity-dependent \conv. Also plotted are the galaxies from the COLDGASS and HRS surveys with well-defined metallicities (as described in the text). Filled symbols and open triangles denote detected sources and 3$\sigma$ upper limits, respectively. The black dotted line shows a linear regression fit, estimated using a Bayesian MCMC technique to account for the upper limits. The green shaded region shows the semi-analytic model prediction from Lagos et al. (2011) (the upper and lower bounds of the coloured region mark the 90th and 10th percentiles of the model distribution).}
\label{fig:h2hi_I97}
\end{figure}
%%%%%%%%%%%%%%%%%%%%

%With $a_1 = -0.56$ and $a_2 = 20.4$ for virial CO line widths (these parameters are essentially independent of UV radiation field strength). 

In addition, we will also discuss the effect on our results of using some additional recent metallicity-dependent conversion factors: those presented by \cite{1997A&A...328..471I}, \cite{2011MNRAS.412..337G},  \cite{2012ApJ...747..124F}, and \cite{2012MNRAS.421.3127N}. Over the metallicity range occupied by our sample (12+log(O/H) $>$ 8.5), most common metallicity-dependent conversion factors exhibit relatively small differences -- the two `outlier' conversion factor prescriptions, predicting the most extreme values of \conv, are \cite{2012MNRAS.421.3127N} (which predicts the lowest values of \conv), and  \cite{1997A&A...328..471I} (which predicts the steepest metallicity dependence, and produces values of \conv\ which trace the upper envelope of measured values for galaxies with sub-solar metallicities). Fig. 9 in  \cite{2013ARA&A..51..207B} shows the relation between metallicity and \conv\ for a range of conversion factors, including \cite{2010ApJ...716.1191W}. Any of our conclusions which are significantly altered by the use of one particular conversion factor will be highlighted below. 
 
Table B3 lists derived gas masses, and other derived properties (including stellar masses, SFRs, and metallicities) for galaxies in ALLSMOG sample. 

\subsection{Comparison to other samples}

Throughout this work, we compare the ALLSMOG sample to two other large surveys for molecular gas in the local Universe -- COLDGASS and the Herschel Reference Survey (HRS). COLDGASS selects galaxies from the GASS survey with stellar masses $>10^{10}$ M$_{\sun}$, while the HRS is a volume-limited sample of galaxies with distances $15 < {\rm D  [Mpc]} < 25$. Both samples have available optical spectra: COLDGASS galaxies are taken from the SDSS spectroscopic survey, and optical spectroscopy for the HRS was published by \cite{2013A&A...550A.114B}. For each of these samples, we also calculate metallicities identically as for ALLSMOG -- the adopted metallicities are taken to be the mean of the two values calculated using the R23 parameter and the [N{\sc ii}]/H$\alpha$ ratio. As before, any galaxies exhibiting a discrepancy of $>0.2$ dex between the two methods were deemed to have unreliable metallicity measurements, and were not used in this analysis. 

We have calculated H$_2$ masses for these samples using the same method as for the ALLSMOG galaxies -- by multiplying their observed CO luminosities by \conv\ (both constant, and metallicity-dependent).

COLDGASS galaxies all appear in the SDSS spectroscopic survey, and -- like ALLSMOG galaxies -- have stellar masses and star formation rates available in the MPA-JHU catalogue. The HRS, however, contains very local galaxies which do not appear in the MPA-JHU catalogue. Stellar masses for the HRS galaxies are available in \cite{2012A&A...544A.101C}. We calculate star formation rates for the HRS galaxies using a combination of UV fluxes (from GALEX, published by \citealt{2012A&A...544A.101C}) and IR fluxes (from the Spitzer Multi-Band Imaging Photometer (MIPS), published by \citealt{2012MNRAS.423..197B}) . We calculate a FUV attenuation using the `IRX' parameter,  IRX$ = \log({\rm L}_{\rm IR}/{\rm L}_{\rm FUV, obs})$:

\begin{equation}
{\rm A(FUV)} = -0.028X^3 + 0.392X^2 + 1.094X + 0.546,
\end{equation}

where $X$ is the IRX parameter \citep{2005MNRAS.360.1413B}. Star formation rates are then calculated from the extinction-corrected FUV luminosity, using

\begin{equation}
\log{\rm SFR} = \log {\rm L}_{\rm FUV, corr} - 9.68,
\end{equation}

which is a SFR prescription calculated specifically for the GALEX bands by \cite{2006ApJS..164...38I}, which we have modified by a factor of 1.5 to convert from their \cite{1955ApJ...121..161S} IMF to the \cite{2001MNRAS.322..231K} IMF used to calculate SDSS parameters.

\section{Results and Discussion}

\subsection{The H$_2$/HI ratio} 
\label{sec:h2hi}

%%%%%%%% FIG 3 %%%%%%%
\begin{figure*}
\centering
\mbox
{
  \includegraphics[width=13cm]{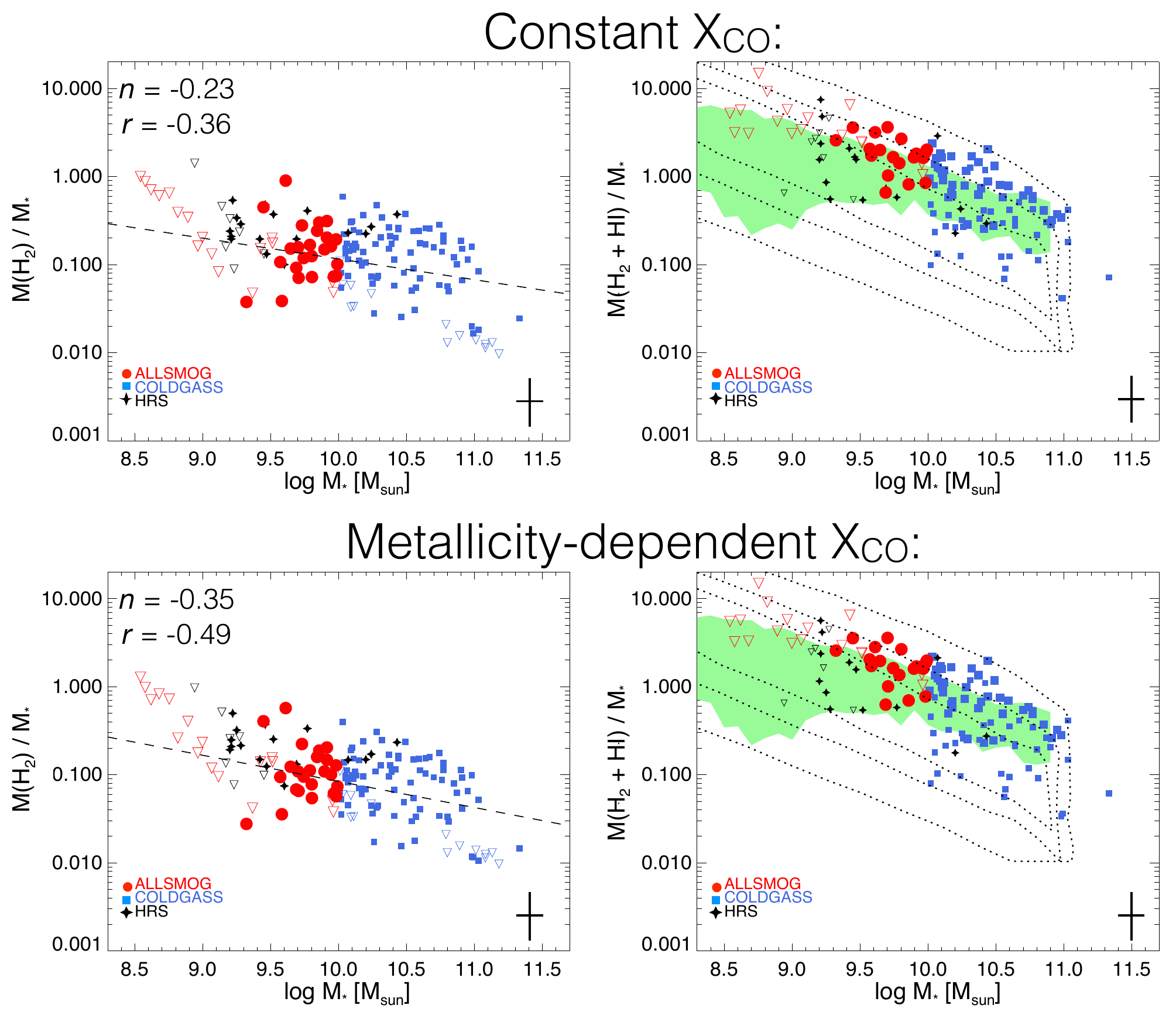}
}
\caption{Gas fractions, plotted against stellar mass, for the ALLSMOG sample. Also plotted are the galaxies from the COLDGASS and HRS surveys with well-defined metallicities (as described in the text). Galaxies with upper limits on their gas mass are plotted with open triangles. The upper row of panels has molecular gas masses calculated using a constant, Milky-Way $\alpha_{\rm CO}$. The lower panels have molecular gas masses calculated using the metallicity-dependent $\alpha_{\rm CO}$ of Wolfire. (2010). The left and right hand panels show (respectively), the molecular, and total gas fraction. Black dashed lines show linear regression fits to all three samples, estimated using a Bayesian MCMC technique to account for the upper limits. Inset legends give the slope of the fits $n$, and the Pearson correlation coefficients $r$. The green shaded area on the total gas fraction panels shows the predictions from cosmological hydrodynamical simulations (Dav\'e et al. 2013), assuming momentum-conserving winds. Dotted contours on the total gas fraction panels mark $2\sigma, 4\sigma, 6\sigma$ contour levels for the distribution of galaxies in the analytic `gas regulator model' predictions of Peng \& Maiolino (2014).}
\label{fig:gasfrac}
\end{figure*}
%%%%%%%%%%%%%%%%%%%% 

%There is a complex interplay between the molecular and atomic gas reservoirs in the interstellar medium. 
The ratio of molecular-to-atomic gas mass in galaxies is dependent on a range of physical processes, including the interstellar radiation field, the pressure of the ISM, and the abundance of dust (which aids the formation of molecules). The ALLSMOG survey, being a low-mass sample, is relatively \HI\ rich compared to more massive samples (e.g., COLDGASS). When using a constant \conv, we measure a mean H$_2$/\HI\ ratio of 0.11, with a 1$\sigma$ scatter of 0.42 dex. Using a metallicity-dependent \conv, we measure a mean H$_2$/\HI\ ratio of 0.08, with a 1$\sigma$ scatter of 0.41 dex.

Fig. \ref{fig:h2hi} shows the H$_2$/\HI\ mass ratio, plotted against stellar mass for ALLSMOG galaxies, and our two comparison samples, COLDGASS and HRS. For our two main choices of \conv\, we find that the H$_2$/\HI\ ratio increases with stellar mass. 

Here (and in our subsequent analyses below) we use regression analysis to fit linear slopes to the data. We employ the IDL code {\tt LINMIX\_ERR}, developed by \cite{2007ApJ...665.1489K} for this purpose. {\tt LINMIX\_ERR} is a Bayesian linear regression estimator, which uses a Markov-Chain Monte Carlo method to fit datasets containing both upper limits, and detected datapoints with heteroscedastic errors. 

%fit linear slopes to the distributions treating the non-detections in two different ways: firstly, we include non-detections at the value of their $3\sigma$ upper limits. Secondly, we exclude non-detections, and fit only to the detected galaxies. 

Though there is considerable scatter at all stellar masses, simple linear regressions applied to the data suggest positive slopes, at high significance. Linear fits give a slope ($n$), for a constant \conv,  of $n=0.66 \pm 0.08$, while the \cite{2010ApJ...716.1191W} metallicity-dependent \conv\ results in a slightly shallower slope of $n=0.53 \pm 0.08$. Formal correlation coefficients ($r$) imply correlation strengths, for a constant \conv, of $r=0.66 \pm 0.06$ -- using a \cite{2010ApJ...716.1191W} metallicity-dependent factor weakens the correlation slightly, giving  $r=0.58 \pm 0.07$. All of the metallicity-dependent conversion factors produce an increase in the H$_2$/\HI\ ratio, with the exception of the \cite{1997A&A...328..471I} factor, which produces a H$_2$/\HI\ ratio which is flat (or even declining) with increasing stellar mass. This is shown in Fig. \ref{fig:h2hi_I97}.

These linear fits are for illustrative purposes only, demonstrating that an increase in H$_2$/\HI\ ratio with stellar mass exists for most choices of \conv. It is unlikely that the true relation between H$_2$/\HI\ ratio and log(M$_*$) is linear; semi-analytic models predict a non-linear increase in H$_2$/\HI\ ratio with log(M$_*$). We overplot on Fig. \ref{fig:h2hi} (and Fig. \ref{fig:h2hi_I97}) the predicted H$_2$/\HI\ ratio from \cite{2011MNRAS.418.1649L}\footnote{The data shown uses a \cite{Baugh:2005aa} model.}. Both a constant conversion factor, and a \cite{2010ApJ...716.1191W} factor, produce a H$_2$/\HI\ ratio distribution in very close accordance with the semi-analytic model data. In contrast, the approximately flat H$_2$/\HI\ ratio distribution produced by a  \cite{1997A&A...328..471I} \conv\ is in conflict with these model predictions. 

We note, though, that there are other semi-analytic models available: \cite{2011MNRAS.418.1649L} also calculate H$_2$/\HI\ ratios using a \cite{2006MNRAS.370..645B} model (which invokes a different star formation quenching mechanism compared to the \cite{Baugh:2005aa} model used above\footnote{Section 2.2 in Lagos et al (2011) explains the difference between the two models in more detail.}),  which slightly under-predicts the H$_2$/\HI\ ratio relative to the data in our samples, though the increase in H$_2$/\HI\ ratio with stellar mass -- and therefore the tension with the \cite{1997A&A...328..471I} conversion factor -- is preserved. 

\subsection{Gas fractions}

Figure \ref{fig:gasfrac} shows gas fraction (defined here as M$_{\rm gas}$/M$_{*}$) plotted against stellar mass for the ALLSMOG sample (as well as the COLDGASS and HRS comparison samples). The left and right hand panels respectively show the molecular and total gas fractions. In the upper set of panels, the molecular gas mass has been derived using a constant \conv, while the lower set shows the molecular gas as derived using a metallicity-dependent \conv.

In all cases, as expected, the molecular gas fraction decreases with increasing stellar mass. The molecular gas fraction decreases approximately monotonically with stellar mass (though with large scatter) across the entire mass range probed by the three samples, $10^9 \lta {\rm M}_* \lta 10^{11}$. 

%%%%%%% FIG 4 %%%%%%%%%
\begin{figure}
\centering
\mbox
{
  \includegraphics[width=8.5cm]{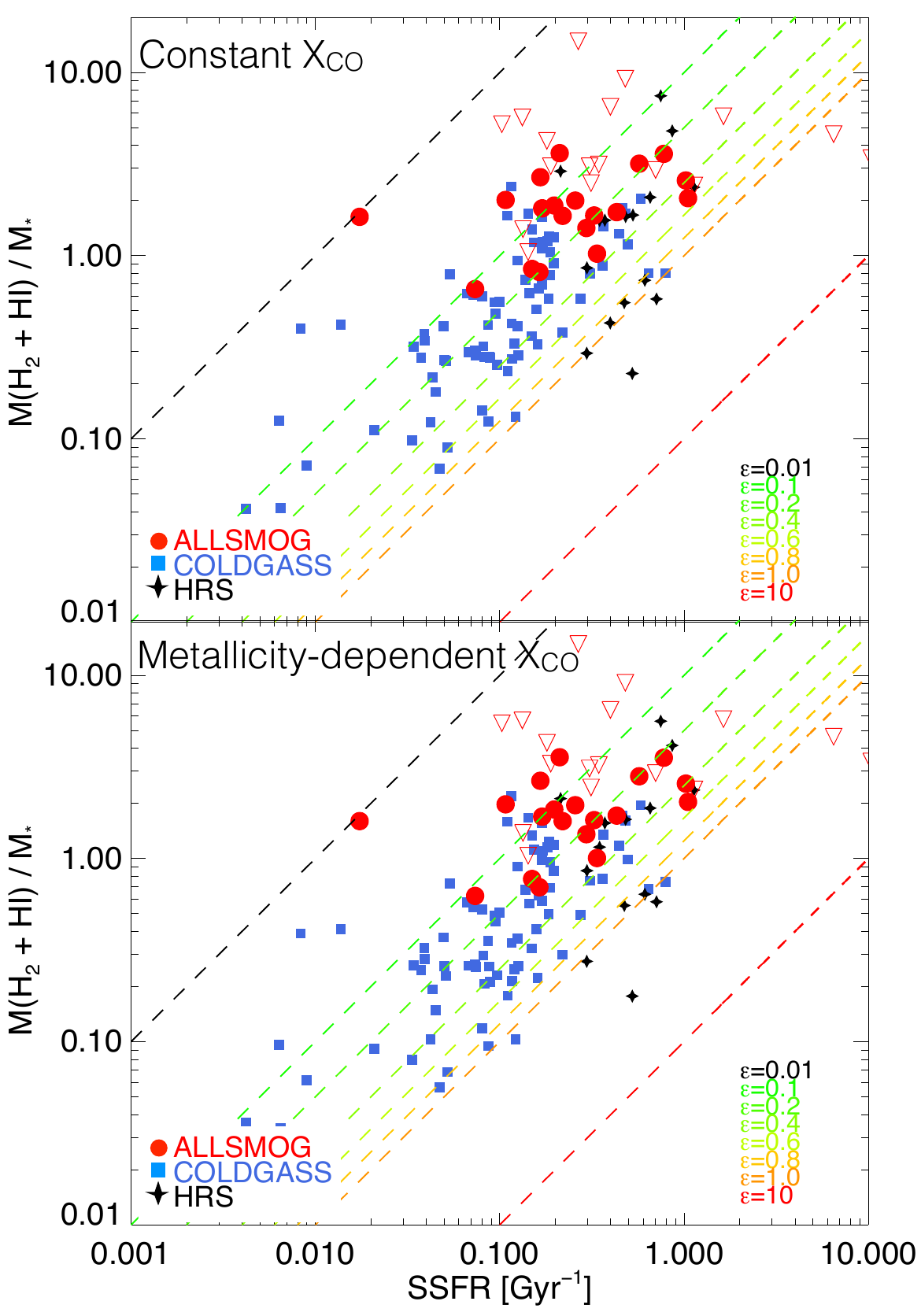}
}
\caption{Gas fractions, plotted against specific star formation rate, for the ALLSMOG sample. Also plotted are the galaxies from the COLDGASS and HRS surveys with well-defined metallicities (as described in the text). The upper panel has molecular gas masses calculated using a constant, Milky-Way appropriate $\alpha_{\rm CO}$. The lower panel has molecular gas masses calculated using the metallicity-dependent $\alpha_{\rm CO}$ of Wolfire et al. (2010). Dotted tracks show different star formation efficiencies ($\epsilon$), ranging from 0.01 to 10 Gyr$^{-1}$ (efficiencies greater than 1.0 are typically seen in starburst systems lying above the `main sequence' of star formation). Star formation efficiencies are here defined in terms of the total (i.e., \HI + H$_2$) gas content.}
\label{fig:gasfrac-ssfr}
\end{figure}
%%%%%%%%%%%%%%%%%%%%

As can be seen in Fig. \ref{fig:gasfrac}, adopting the \cite{2010ApJ...716.1191W} metallicity-dependent \conv\ causes the gas fraction vs. M$_*$ relation to steepen, and become more tightly (anti-) correlated. Linear regressions applied to the molecular gas fraction data produce a slope of $-0.23 \pm 0.06$ for a constant \conv: when adopting the metallicity-dependent conversion factor, the slope steepens to $-0.35 \pm 0.05$. Likewise, the distributions are more tightly (anti-) correlated when using the metallicity-dependent \conv. The Pearson correlation coefficient is $r=-0.36 \pm 0.09$ for a constant \conv, while adopting the metallicity-dependent \conv\ causes the correlation to tighten somewhat, to $r=-0.49 \pm 0.08$. These conclusions remain essentially identical if we use either of the four additional metallicity-dependent \conv\ prescriptions given above: fitting to all galaxies using \cite{2012ApJ...747..124F} gives a slope of $n = -0.36 \pm 0.06$ and $r=-0.5$;  a \cite{2011MNRAS.412..337G} factor gives $n = -0.26 \pm 0.04$ and $r=-0.39$, using \cite{2012MNRAS.421.3127N} gives $n = -0.45 \pm 0.04$ and $r=-0.61$, and using \cite{1997A&A...328..471I} gives by far the steepest value,  $n = -0.86\pm 0.07$ and $r=-0.70$. Across all our samples, we observe a monotonic decrease in molecular gas fraction with increasing stellar mass, which exists independent of the choice of CO/H$_2$ conversion factor. %We do not observe a `levelling off' of the gas fraction at lower stellar masses, as predicted by some models (cite?).

%%%%%%% FIG 5 %%%%%%%%%
\begin{figure*}
\centering
\mbox
{
  \includegraphics[width=13cm]{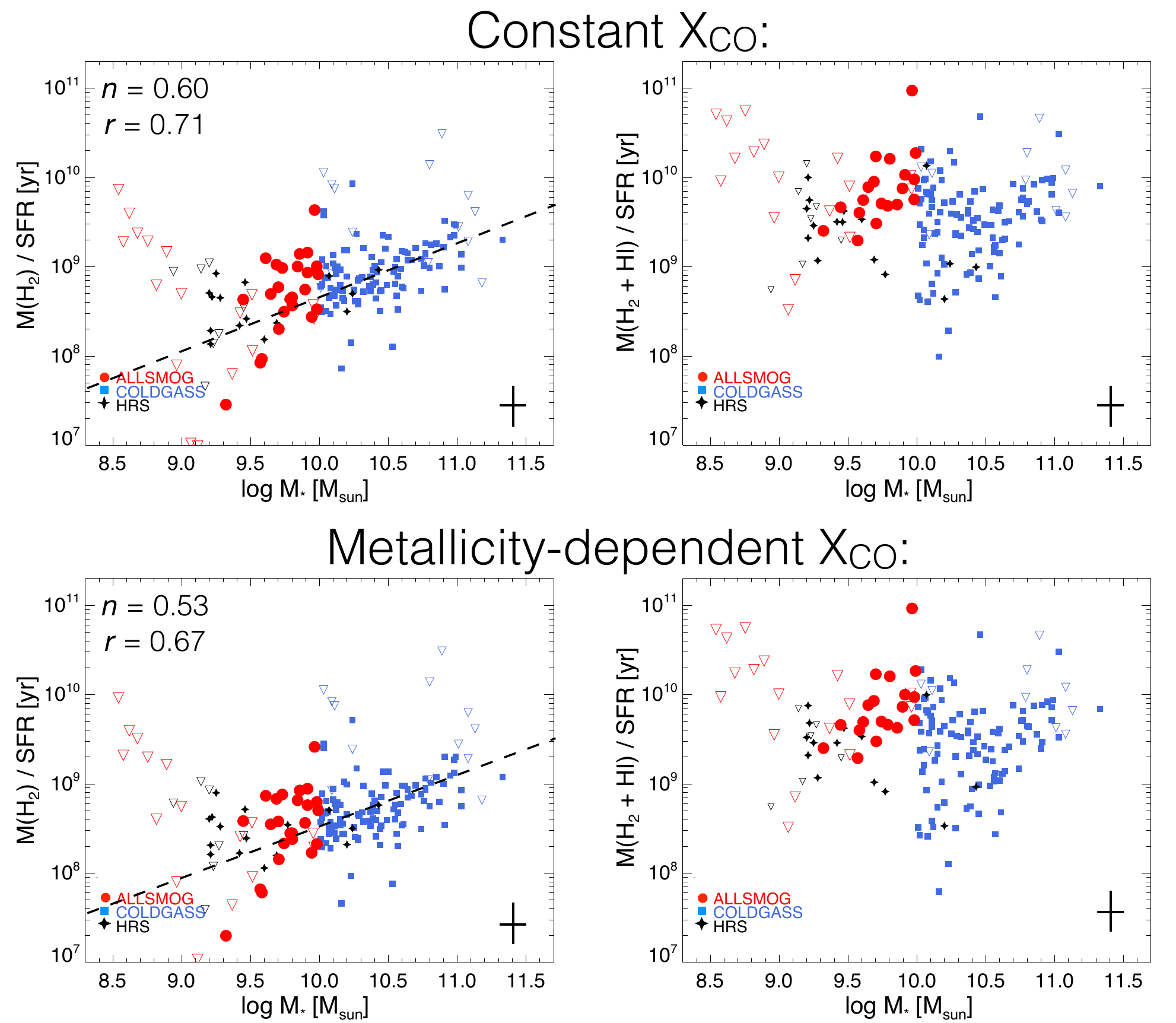}
}
\caption{Gas consumption timescales, plotted against stellar mass, for the ALLSMOG sample. Also plotted are the galaxies from the COLDGASS and HRS surveys with well-defined metallicities (as described in the text). Filled symbols and open triangles denote detected sources and 3$\sigma$ upper limits, respectively. The upper row of panels has molecular gas masses calculated using a constant, Milky-Way appropriate $\alpha_{\rm CO}$. The lower panels have molecular gas masses calculated using the metallicity-dependent $\alpha_{\rm CO}$ of Wolfire et al. (2010). The left and right hand panels show (respectively), the molecular, and total gas fraction. There is a positive correlation between molecular gas consumption time and stellar mass, regardless of the choice of conversion factor. Linear regression fits to the molecular gas timescales are shown in the left panels: the slopes are $0.63 \pm 0.06$ for the constant \conv, and $0.48\pm0.06$ for the metallicity-dependent \conv. The inset caption gives the slope of the linear fit ($n$), and the Pearson correlation coefficient ($r$).}
\label{fig:tau}
\end{figure*}
%%%%%%%%%%%%%%%%%%%%

The right hand panels of Fig. \ref{fig:gasfrac} show the {\it total} gas fraction, where `total gas' is defined as M(H$_2$) + M(\HI) (a 36\% correction to account for helium is included in both of the CO/H$_2$ conversion factors) plotted against stellar mass. Due to the abundance of atomic gas in lower-mass systems (see \S5.1 above), total gas fractions are, in general, far higher than the molecular fractions. We observe essentially the same trend, however: a monotonic decrease in total gas fraction across the entire stellar mass range of our samples. The mean total gas fraction for the ALLSMOG sample is $1.85$.%, with a $1 \sigma$ rms of $\sim 0.2$ dex. 

It is difficult to quantify any trends in total gas fraction between the three samples, unfortunately, due to differing sample selection effects. While the ALLSMOG sample selected \HI\ observations from publicly-available catalogs, the COLDGASS survey (as a subset of the GASS survey) was designed to observe deeper in \HI\ than many other wide-field \HI\ surveys. As a result, there is a clear selection effect-induced discontinuity in the distribution around ${\rm M}* = 10^{10} {\rm M}_{\sun}$, with a subset of the more massive COLDGASS sample having far deeper \HI\ data (and exhibiting lower total gas fractions) than ALLSMOG/HRS galaxies with ${\rm M}* < 10^{10} {\rm M}_{\sun}$. As a result, we compare our gas fraction trends to simulations, but refrain from quantifying trends for the samples taken as a whole.

We have plotted on the right-hand panels of Fig. \ref{fig:gasfrac} gas fraction predictions from cosmological hydrodynamical simulations taken from \cite{2013MNRAS.434.2645D}. We plot the model prediction which assumes `hybrid energy/momentum-driven winds' winds (ezw in their nomenclature). We choose the compare our data to this `ezw' model, as \cite{2013MNRAS.434.2645D} find this model matches observables (such as the \HI\ mass function) better than their other wind models; for more details we refer readers to \S2.1 of \cite{2013MNRAS.434.2645D}. 

%In both the constant \conv\ and metallicity-dependent \conv\ cases, the \cite{2013MNRAS.434.2645D} hydrodynamical model under-predicts gas fractions relative to our data. This issue was noted by \cite{2011MNRAS.416.1354D} to exist at high stellar masses, but at that time there were no surveys for molecular gas in lower mass galaxies available. We find that the extent of the discrepancy between the model and observations increases towards low stellar masses, due to the fact that the hydrodynamical simulations predict a `flattening' in gas fraction at low stellar masses (possibly caused by incorrectly modelled preventative feedback). Our observations suggest no such downturn -- as stellar mass decreases, the total gas fraction rises, to the limits of our detected survey data (M$* \sim 10^{9} {\rm M}_{\sun}$).

In both the constant \conv\ and metallicity-dependent \conv\ cases, the \cite{2013MNRAS.434.2645D} hydrodynamical model predicts total gas fractions roughly in accordance with our data. Observed galaxies in our sample trace the `upper envelope' of the model data, but this is to some extent to be expected (as our use of \HI\ parent surveys provides samples of galaxies which are more \HI\ rich than average).

As well as the hydrodynamic model predictions from \cite{2013MNRAS.434.2645D}, we also plot on Fig. \ref{fig:gasfrac} analytic predictions from the \cite{2014arXiv1402.5964P}  `gas regulator' model. This model (often called the `bathtub' model) can predict galaxy behaviour by analytically solving coupled equations describing a small number of physical parameters -- gas accretion, outflow, star formation and metal production. It can be seen in Fig. \ref{fig:gasfrac} that this analytical model produces a close match to the observed gas fraction data. While the majority of galaxies in all three surveys clearly represent gas-rich systems, the distribution, and in particular the monotonic slope of the gas fraction -- stellar mass relation, is reproduced well by the \cite{2014arXiv1402.5964P} analytic model.

Figure \ref{fig:gasfrac-ssfr} shows the total gas fraction as a function of the specific star formation rate (SSFR = SFR/M$*$). Gas fraction correlates positively with specific star formation rate (with considerable scatter). This scatter can be interpreted in terms of star formation efficiency variations, whereby galaxies at a given gas fraction with higher star formation efficiencies have higher specific star formation rates (though the scatter in the H$_2$/\HI\ ratio at a given stellar mass will contribute, due to the close dependence between SFR and H$_2$ content). 

On Fig. \ref{fig:gasfrac-ssfr} we have also shown `tracks' of varying star formation efficiency, calculated using the expression:

\begin{equation}
f_g = \frac{{\rm SSFR}}{\epsilon}
\end{equation}

where $f_g$ is the total gas-to-stellar mass ratio, and $\epsilon$ is the star formation efficiency, SFR/M(gas). This expression follows analytically from the definitions of SSFR and gas fraction. 

It can be seen that the vast majority of the galaxies, across all three samples, lie within regions with implied star formation efficiencies of $0.1 < \epsilon/ {\rm Gyr}^{-1}< 1.0$. There are a few outliers with higher implied efficiencies, which are likely to be  systems lying above the main sequence, undergoing a starburst event (caused by an interaction, or just stochastically clumpy gas accretion). Galaxies in the ALLSMOG sample have typical star formation efficiencies of $\epsilon <0.5 \, {\rm (Gyr)}^{-1}$. No galaxies in the ALLSMOG sample -- even those not detected in CO emission -- imply star formation efficiencies $<0.01 \, {\rm (Gyr)}^{-1} $.

\subsection{Gas consumption timescales}

It is also possible to interpret the ratio between gas mass and SFR in terms of a gas consumption timescale, defined here as $\tau_{\rm gas} =  {\rm M}_{\rm gas}$/SFR ($= 1/\epsilon$). 

Figure \ref{fig:tau} shows the gas consumption timescales for the ALLSMOG sample (as well as the comparison samples, COLDGASS and the HRS). Fig. \ref{fig:tau}  is structured following Fig. \ref{fig:gasfrac} -- the left and right hand panels show, respectively, the molecular and total gas consumption timescales, while the upper and lower panels calculate molecular gas masses using a constant \conv\ (upper panels), and a \cite{2010ApJ...716.1191W} metallicity-dependent \conv\ (lower panels). The mean molecular gas consumption timescale for the ALLSMOG galaxies detected in CO is $0.78 \pm 0.1$ Gyr for a constant \conv\, and $0.50 \pm 0.1$ Gyr for the metallicity-dependent \conv. The rms scatter in $\log (\tau_{\rm H2})$ is 0.44 dex.

We confirm the trends found by both \cite{2011MNRAS.415...61S} and \cite{2014arXiv1401.8101B}: that the molecular gas consumption timescale  $\tau_{\rm H2}$ is not constant, but increases with stellar mass over the full mass range of the detected samples (M$_* \simgt 10^9 {\rm M}_{\sun}$). Molecular gas depletion timescales vary from $\sim$2 Gyr at the most massive end of the distribution inhabited by COLDGASS galaxies (M$_* > 10^{11} {\rm M}_{\sun}$), to $\sim$100 Myr for the lowest mass detected galaxies (M$_* \sim 10^{9} {\rm M}_{\sun}$).

However, in addition to this finding (and in opposition to the results of \citealt{2014arXiv1401.8101B}), we find that this positive trend persists, even when using a non-constant CO/H$_2$ conversion factor. Again, applying the linear regression analysis discussed above, we find that when using a constant \conv, $\tau_{\rm H2}$ increases with stellar mass with a slope of $0.60 \pm 0.06$ (and a Pearson correlation coefficient of $r=0.71$). Employing the \cite{2010ApJ...716.1191W} metallicity-dependent \conv\ flattens the relation slightly, but the correlation still persists, with a slope of $0.53 \pm 0.07$ (and $r=0.67$). Indeed, we find this positive correlation between gas consumption timescale and stellar mass persists at high significance when using three of the four additional metallicity-dependent \conv\ prescriptions listed in \S\ref{sec:h2} above: a \cite{2012ApJ...747..124F} conversion factor results in a slope of index $0.46\pm0.07$ ($r=0.61$); using \cite{2011MNRAS.412..337G} produces a slope of index $0.59\pm0.06$ $(r=0.70)$; and \cite{2012MNRAS.421.3127N} produces a slope of $0.41 \pm0.07$ ($r=0.56$).

In contrast to these findings, use of the \cite{1997A&A...328..471I} conversion factor produces an essentially flat relationship between molecular gas consumption timescale and stellar mass  (albeit with nearly an order of magnitude scatter), with a slope of $-0.11 \pm 0.10$ (and $r=-0.13$). This (close to) flat relationship would imply an approximately constant molecular gas consumption timescale of $\log \tau_{\rm H2} = 8.25 \pm 0.71$ yr across all three samples.

While the use of a constant, Milky-Way value for \conv\ produces the strongest (and steepest) positive correlation between gas consumption timescale and stellar mass, the correlation, whereby the most massive galaxies have molecular gas consumption timescales over an order of magnitude longer than the least massive galaxies observed here, persists for three of the four metallicity-dependent CO/H$_2$ conversion factors used in this work. As above with the H$_2$/\HI\ ratio, the exception to this is the \cite{1997A&A...328..471I} conversion factor, which produces a relation between H$_2$ consumption time and stellar mass which is approximately flat.  

%While use of the remaining conversion factor \citep{1997A&A...328..471I} produces an approximately flat gas consumption time, it also conflicts with the \conv\ derivations for local disk galaxies presented by \cite{2013ApJ...777....5S}\footnote{See \cite{2013ARA&A..51..207B} for discussion.}, and produces results which are in tension with the H$_2$/\HI\ ratio model predictions discussed in \S5.1 above.

The relation between {\it total} gas consumption time and stellar mass is more difficult to interpret from the samples used in this work, due to the \HI\ selection effects mentioned above. Certainly, at stellar masses $<10^{10} \, {\rm M}_{\sun}$ the detected ALLSMOG galaxies show signs of a positive correlation between $\tau_{\rm gas}$ and M$_*$. Due to the differing \HI\ sample selection effects above and below $10^{10} \, {\rm M}_{\sun}$, however, it is difficult to identify any trends existing across the full mass range. It is likely that deep dedicated \HI\ observations at low stellar masses are required in order to fully examine trends in total gas consumption timescale.

\section{Summary}

We have presented the initial data release and analysis from the ALLSMOG survey, an public legacy survey for molecular gas in local low-mass galaxies. We have used APEX to observe the CO($2-1$) emission line in 42 galaxies with stellar masses in the range 8.5 $<$ log(M$*$/M$_{\sun}$) $<$ 10, redshifts in the range $0.01 < z < 0.03$, and metallicities $12 + \log({\rm O/H}) > 8.5$. We detect a total of 25 of 42 galaxies observed -- a detection rate of 60\%. There is a strong stellar mass bias in the detection statistics, with all galaxies with stellar masses ${\rm M}_*< 10^{9} {\rm M}_{\sun}$ (with one exception) being undetected.

We use our metallicities, derived using optical strong line methods, to calculate metallicity-based CO/H$_2$ conversion factors, avoiding the unphysical assumption of a constant \conv. Our main conclusions are as follows:

\begin{enumerate}

\item{The ratio  M(H$_2$)/M(\HI) increases with stellar mass, albeit with large scatter, in agreement with both semi-analytic predictions and previous work. Importantly, this result remains robust when using a constant CO/H$_2$ conversion factor, and three of the four metallicity-dependent CO/H$_2$ conversion factors used in this work. The exception -- the  \cite{1997A&A...328..471I} \conv\ -- produces a M(H$_2$)/M(\HI) ratio which is roughly flat with stellar mass, in conflict with model predictions.}\\

\item{Molecular gas fractions decrease with increasing stellar mass -- this conclusion holds independent of the choice of CO/H$_2$ conversion factor. Total gas fractions (H$_2$+\HI+He) also decrease with stellar mass, in accordance with predictions from both analytic and hydrodynamical models}\\

%-- however, our total gas fractions at low stellar mass are typically $\sim$1 dex higher than model gas fractions seen in hydrodynamical simulations. Some of this difference may be attributed to the shallow \HI\ data available, but it is likely that there remains a tension between hydrodynamic and observed gas fractions which extends down to at least $10^9 {\rm M}_{\sun}$. The analytic gas fraction predictions given in \cite{2014arXiv1402.5964P} are a good match to the data. }\\

\item{We confirm the non-universality of the molecular gas consumption timescale across 2 orders of magnitude in stellar mass, finding depletion timescales varying monotonically from $\sim$2 Gyr at the massive end of the distribution (inhabited by COLDGASS galaxies)  to $\sim$100 Myr for the lowest mass galaxies in our survey. Again, we find that this result persists when using a constant conversion factor, and three of the four CO/H$_2$ conversion factor prescriptions used this this work. Use of the \cite{1997A&A...328..471I} conversion factor produces a molecular gas consumption time which is approximately constant.}\\

%\item{When examining both the M(H$_2$)/M(\HI) ratio, and the molecular gas consumption timescale, use of the \cite{1997A&A...328..471I} conversion factor produces flat relations which are in opposition to trends found using both a range of other metallicity-dependent conversion factors, and a constant, Milky Way-appropriate factor.}

\end{enumerate}

\section*{Acknowledgments}

This publication is based on data acquired with the Atacama Pathfinder Experiment (APEX), with ESO programme ID number 192.A-0359. APEX is a collaboration between the Max-Planck-Institut f{\"u}r Radioastronomie, the European Southern Observatory, and the Onsala Space Observatory. We would like to extedt our thanks to the staff of the APEX observatory for their support while observing these data, and to the anonymous referee whose comments helped improve this manuscript. This research has made use of NASA's Astrophysics Data System. This work was co-funded under the Marie Curie Actions of the European Commission (FP7-COFUND).

%\bibliography{/Users/Matt/Documents/mybib}{}
\bibliography{/Users/matthew/Documents/mybib}{}
\bibliographystyle{mn2e}

\appendix

\section{Aperture correction calculations}
\label{sec:aper}

%In most cases, the extent of the galaxy is such that the 27$''$ APEX beam will succeed in covering the majority of the CO emitting area. 

In order to derive total gas masses it is necessary to derive aperture correction factors, to correct for any potential flux lying outside the beam emitting area.

To account for this possible source of uncertainty, we calculate for each ALLSMOG galaxy a `beam coverage correction', which will allow us to correct for potential missing flux due to the beam being smaller than the CO emitting region. B-band sizes (in terms of r$_{25}$, the 25th magnitude isophote) are available for each ALLSMOG galaxy, and are listed in Table B1. We calculate predicted CO exponential disc scaling lengths, $h$, using the relation derived from resolved CO observations (\citealt{1995ApJS...98..219Y}; \citealt{2009AJ....137.4670L}), $h_{CO(2-1)} = 0.2\pm0.05 r_{25}$.

An exponential disc has a radial surface brightness profile given by 

\begin{equation}
I(R) = I(0) e^{(-R/h)},
\end{equation}

where $I(0)$ is the central brightness (irrelevant here, as we are simply interested in the percentage of the total flux recovered by the beam), $R$ is the radius, and $h$ is the scaling length. 

For each galaxy, we integrate its exponential disc model out to the radius covered by the 27$''$ beam. However, inclination effects must also be taken into account. For a fully face-on galaxy, the fraction of flux falling within the beam will be given by integrating a 2D exponential disc model out to the beam radius (relative to the total flux):

\begin{equation}
{\rm Flux \; fraction} = \frac{\int_0^{2\pi}\int_0^{\rm Beam} e^{-R/h} R\, dr\,d\theta}{\int_0^{2\pi}\int_0^{\rm \infty} e^{-R/h} R\, dr\,d\theta}
\label{eq:inc_a}
\end{equation}

At higher inclination angles, more CO flux will fall inside the beam, and the solution will tend towards a 1D exponential model (for a fully edge-on galaxy):

\begin{equation}
{\rm Flux \; fraction} = \frac{\int_0^{\rm Beam} e^{-R/h} dR}{\int_0^{\rm \infty} e^{-R/h} dR}
\label{eq:inc_b}
\end{equation}

We therefore calculate a flux correction for each ALLSMOG galaxy by taking a linear combination of these two extreme situations, with the relative contribution from each given by the inclination angle $i$:

\begin{multline}
F = (\sin i) \left(\frac{\int_0^{\rm Beam} e^{-R/h} dR}{\int_0^{\rm \infty} e^{-R/h} dR} \right)  \\ + (1-\sin i)  \left( \frac{\int_0^{2\pi}\int_0^{\rm Beam} e^{-R/h} R\, dR\,d\theta}{\int_0^{2\pi}\int_0^{\rm \infty} e^{-R/h} R\, dR\,d\theta} \right)
\label{eq:inc}
\end{multline}

It can be seen that at $i=0^{\circ}$ and $i=90^{\circ}$, this expression becomes the face-on and edge-on solutions (respectively). 

\begin{figure*}
\centering
\mbox
{
  \includegraphics[width=17cm]{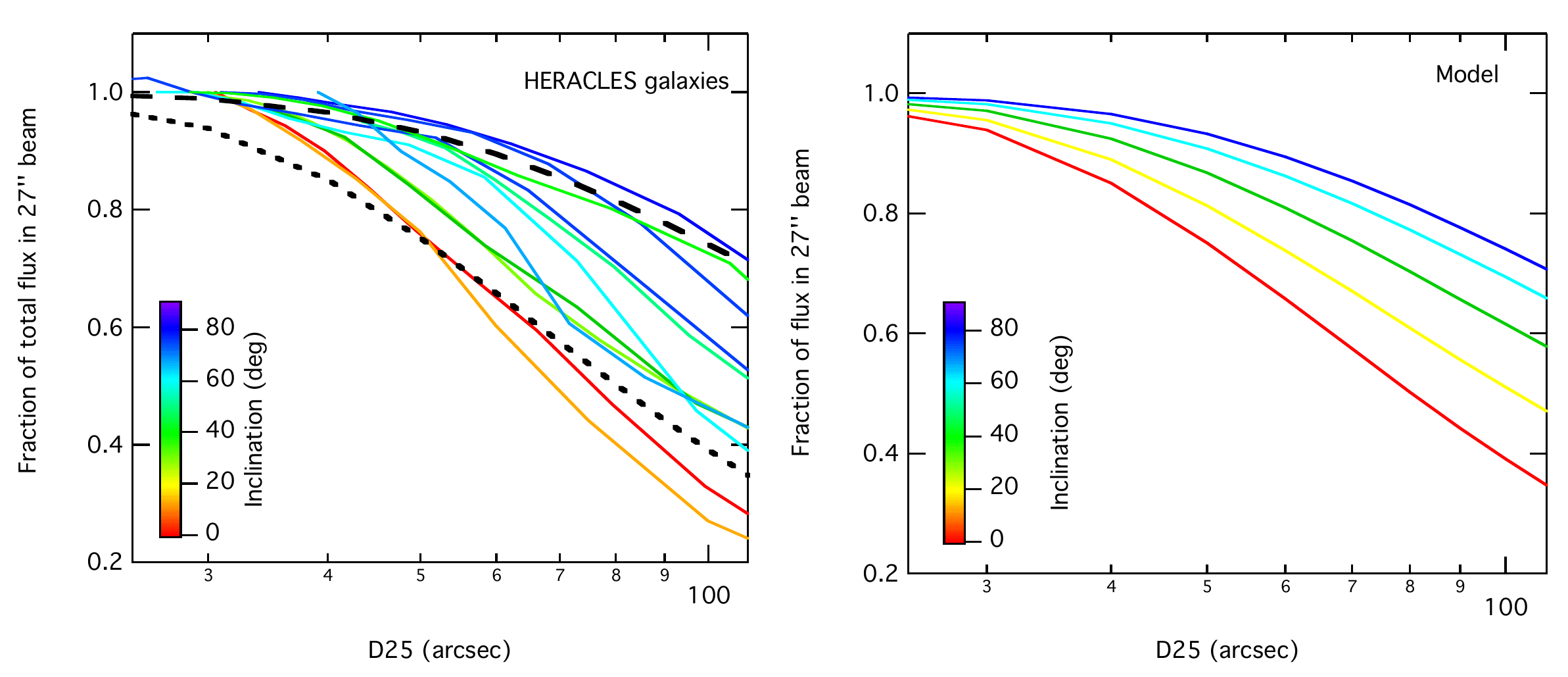}
}
\caption{Plots showing the fraction of total CO$(2-1)$ flux is recovered by a $27''$ beam, plotted against the optical size of the galaxy. The left panel shows empirically measured flux fractions, taken from maps of the HERACLES sample (Leroy et al. 2011) where we have manually redshifted each map to simulate a range of sizes. The coloured lines show individual galaxies, while the black dotted and dashed lines show the 2D and 1D exponential solutions, respectively. While it is clear that the flux fraction drops as a function of optical size, there is also an inclination effect (by which at a given optical size, a face-on galaxy will have more of its flux outside the beam than an edge-on galaxy). As expected, 1D and 2D analytical solutions effectively define the boundaries of the distribution, being good matches for the most edge on and face on galaxies, respectively. The right panel shows the predicted flux fractions, calculated using model given in Eq. \ref{eq:inc}. }
\label{fig:inc}
\end{figure*}

We have empirically tested this technique using resolved CO($2-1$) maps of the HERACLES sample \citep{2009AJ....137.4670L}. We manually redshifted each HERACLES CO($2-1$) map, so that their optical diameters spanned a range $25 < $ D25$'' < 100$ (approximately the range of diameters in the ALLSMOG sample). For each HERACLES galaxy, we then calculated how much of the total CO$(2-1)$ flux would be recovered by a $27''$ beam as a function of its size. The results are shown in Fig. \ref{fig:inc} (left). There is a clear inclination effect (at a given optical size, a face-on galaxy will have more of its flux outside the beam than an edge-on galaxy). We have also plotted on Fig. \ref{fig:inc} (left) the analytic solutions found by integrating 1D and 2D exponential functions (Equations \ref{eq:inc_a} and \ref{eq:inc_b}). As expected, the two cases define the approximate boundaries to the distribution, with the 2D solution being a good model for the most face-on galaxies, and the 1D solution being a good model for the most edge-on galaxies. 

The right panel of Fig. \ref{fig:inc} shows the flux fraction for various inclinations, as predicted by Eq. \ref{eq:inc}. There is a good match between the model and the HERACLES data, and as such we adopt this method for calculating aperture corrections for the ALLSMOG sample. In the four cases where no inclination angle is listed for an ALLSMOG galaxy, we adopt an inclination-averaged angle of $< \sin i > = \pi/4$ ($\theta \sim 51.7$ -- see appendix A of \cite{2009ApJ...697.2057L} for the derivation of this value). 

Table B2 lists the percentage of the total flux recovered by our observations, calculated using Eq. A4. All CO luminosities and H$_2$ gas masses used throughout this work (and listed in Table AB) are calculated with this beam coverage correction applied.

%%%%%%%% FIG 2 %%%%%%%
%%%%%%%%%%%%%%%%%%%%

  \newpage

\section{Data tables}

\include{tab1}
\include{tab2}
\include{tab3}

\section{Maps and Spectra}

\begin{figure*}
\centering
\mbox
{
  \includegraphics[width=13cm, clip=true, trim=5cm 3cm 10cm 1cm]{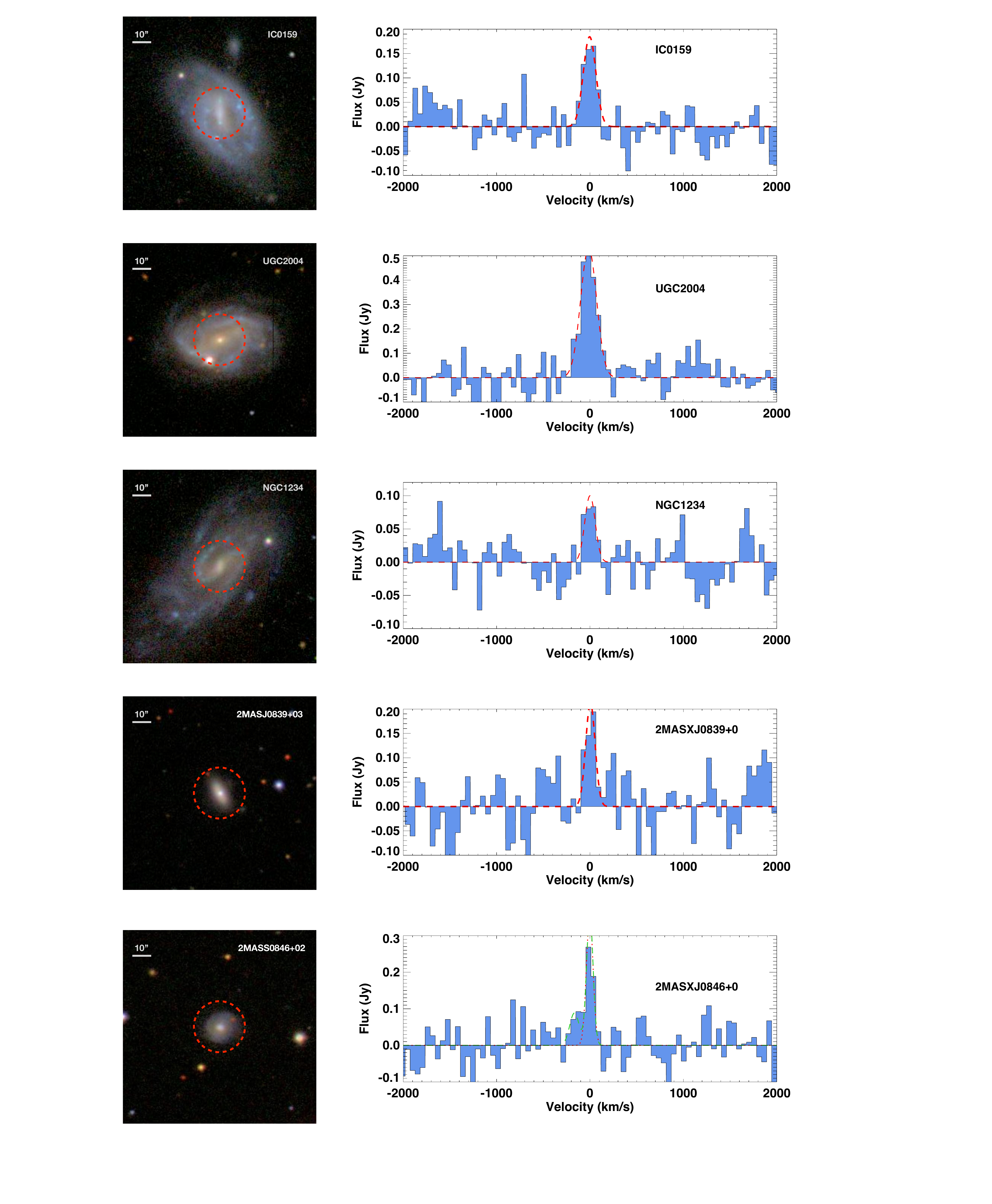}
}
\caption{SDSS cutout images of galaxies in the ALLSMOG sample detected in CO($2-1$) emission, and their corresponding spectra. Galaxies are shown at a constant size scale ($10''$ is shown inset in the upper left). In each left hand panel, the central circle shows the size of the $27''$ APEX beam. Spectra are presented at 60 km\,s$^{-1}$ resolution, and overlaid with the best fitting Gaussian profile.}
\label{fig:mos1}
\end{figure*}

\begin{figure*}
\centering
\mbox
{
  \includegraphics[width=13cm, clip=true, trim=5cm 3cm 10cm 1cm]{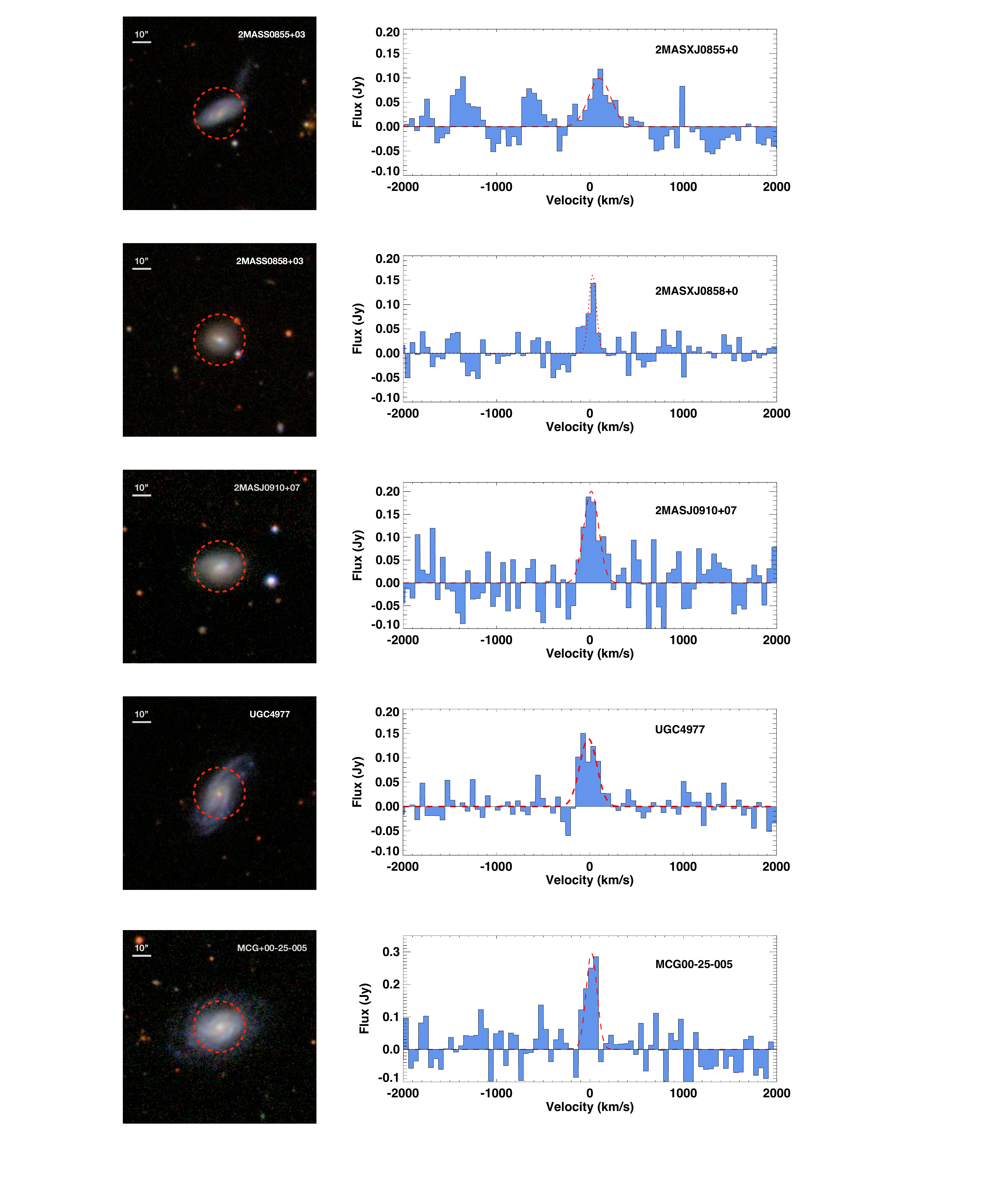}
}
\caption{Panels are the same as Fig. \ref{fig:mos1}. The galaxy MCG+00-27-013 exhibits a double-peaked spectrum, and has been fitted with both a single and a double Gaussian profile (as discussed in the text).}
\label{fig:mos2}
\end{figure*}

\begin{figure*}
\centering
\mbox
{
  \includegraphics[width=13cm, clip=true, trim=5cm 3cm 10cm 1cm]{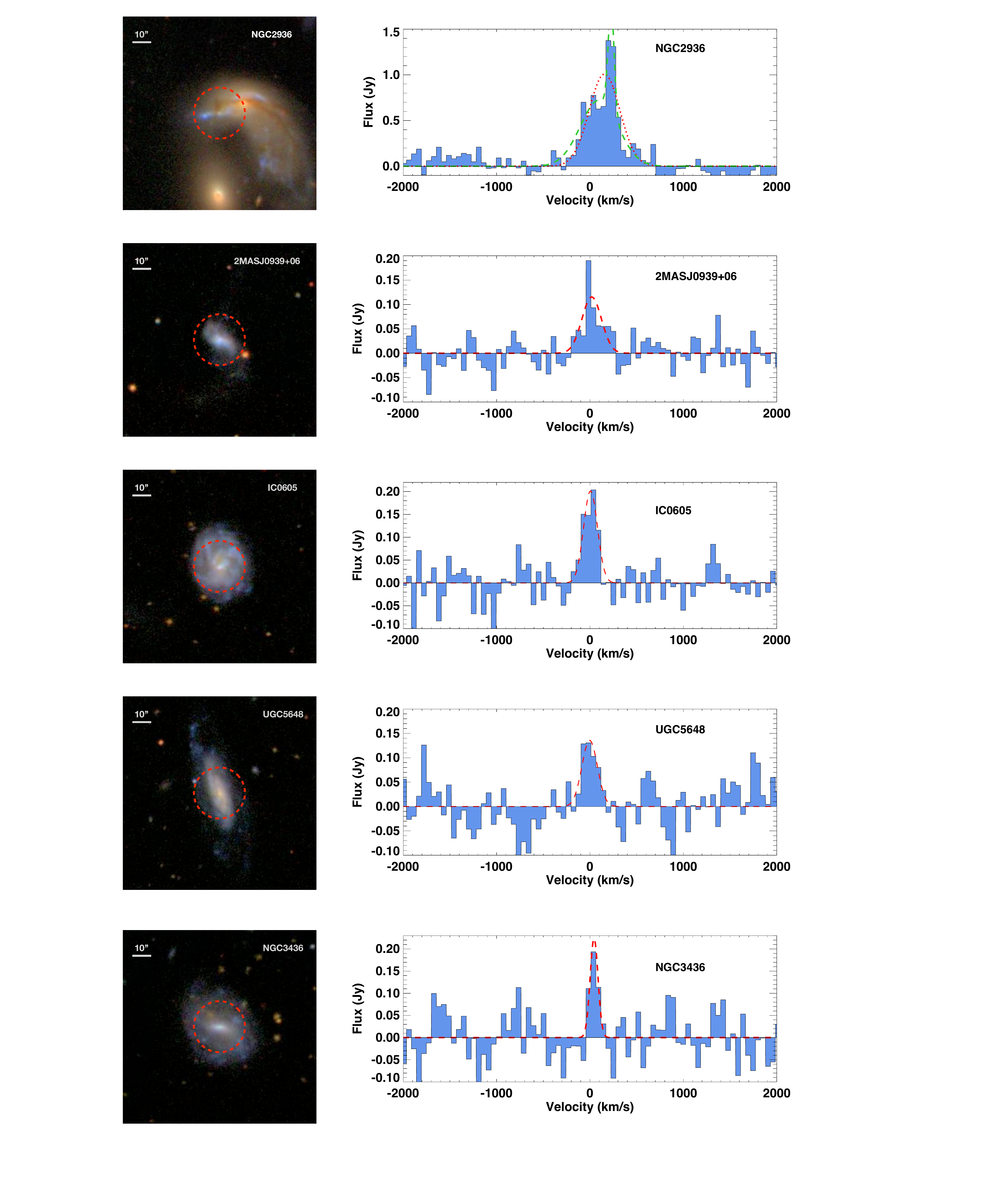}
}
\caption{Panels are the same as Fig. \ref{fig:mos1}. The galaxy UGC06838 has been presented at higher spectral resolution (20 km\,s$^{-1}$ channels) due to the narrow linewidth.}
\label{fig:mos3}
\end{figure*}

\begin{figure*}
\centering
\mbox
{
  \includegraphics[width=17cm]{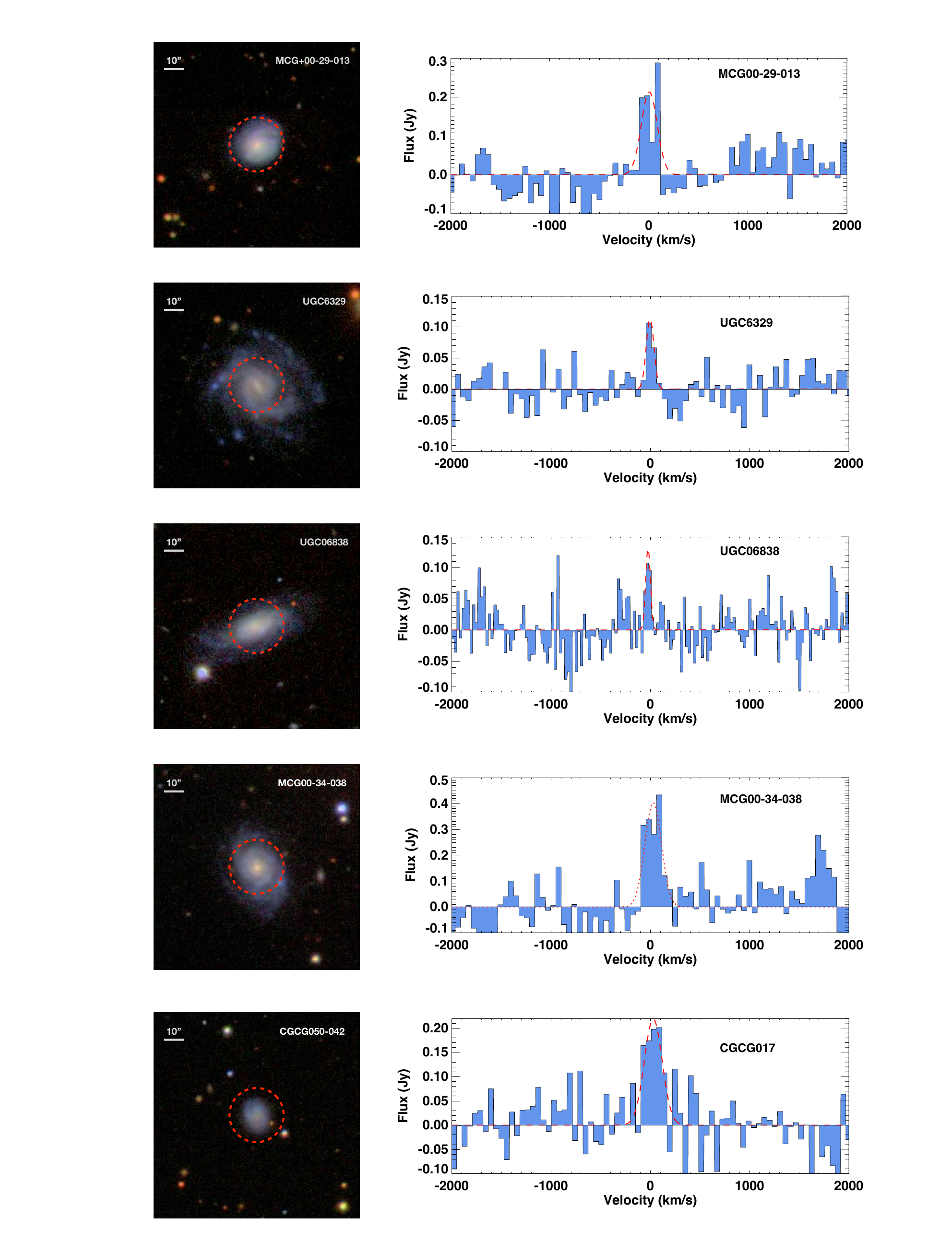}
}
\caption{Panels are the same as Fig. \ref{fig:mos1}.}
\label{fig:mos4}
\end{figure*}

\begin{figure*}
\centering
\mbox
{
  \includegraphics[width=13cm, clip=true, trim=5cm 3cm 10cm 1cm]{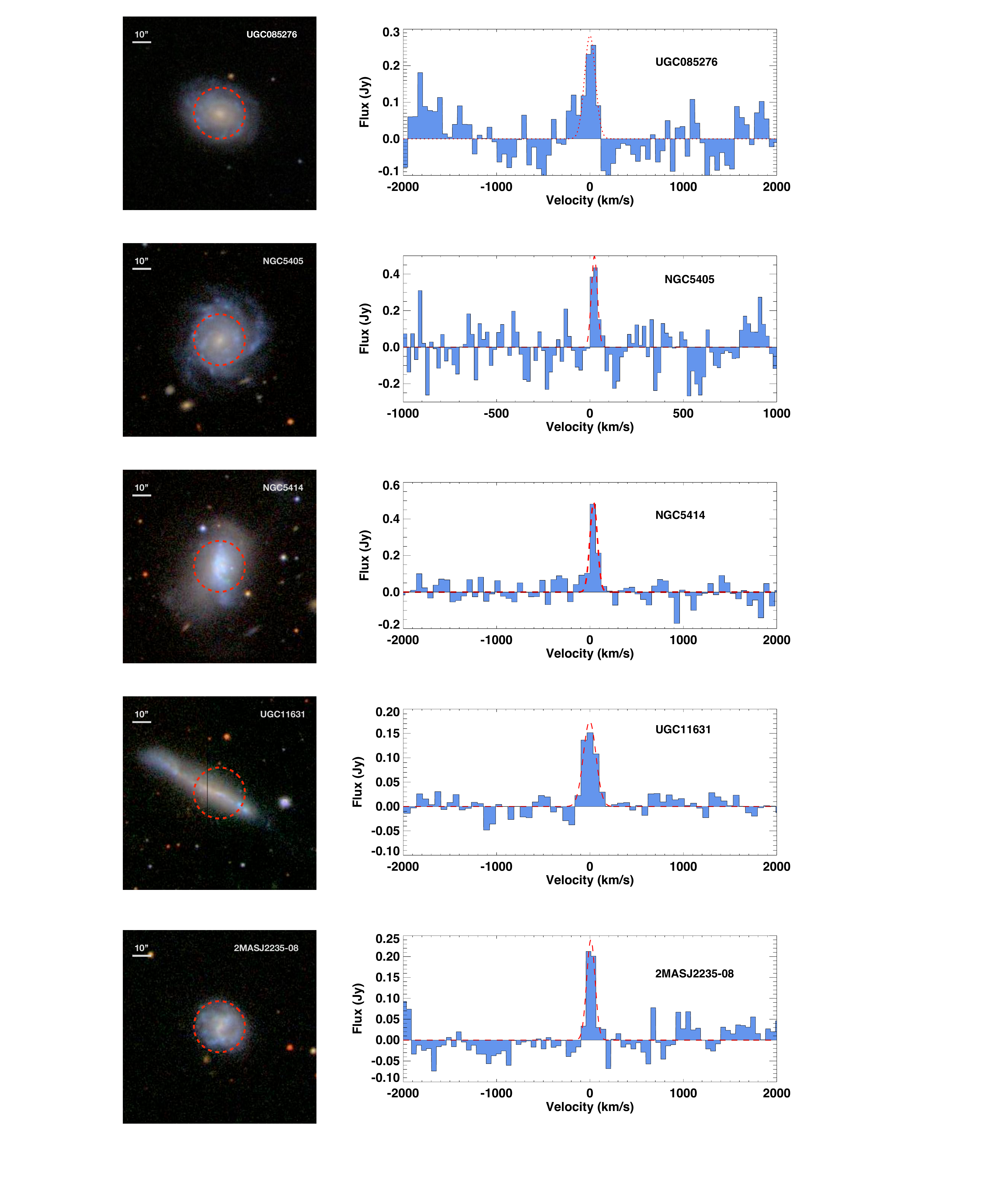}
}
\caption{Panels are the same as Fig. \ref{fig:mos1}.}
\label{fig:mos4}
\end{figure*}

\begin{figure*}
\centering
\mbox
{
  \includegraphics[width=13cm, clip=true, trim=5cm 3cm 10cm 1cm]{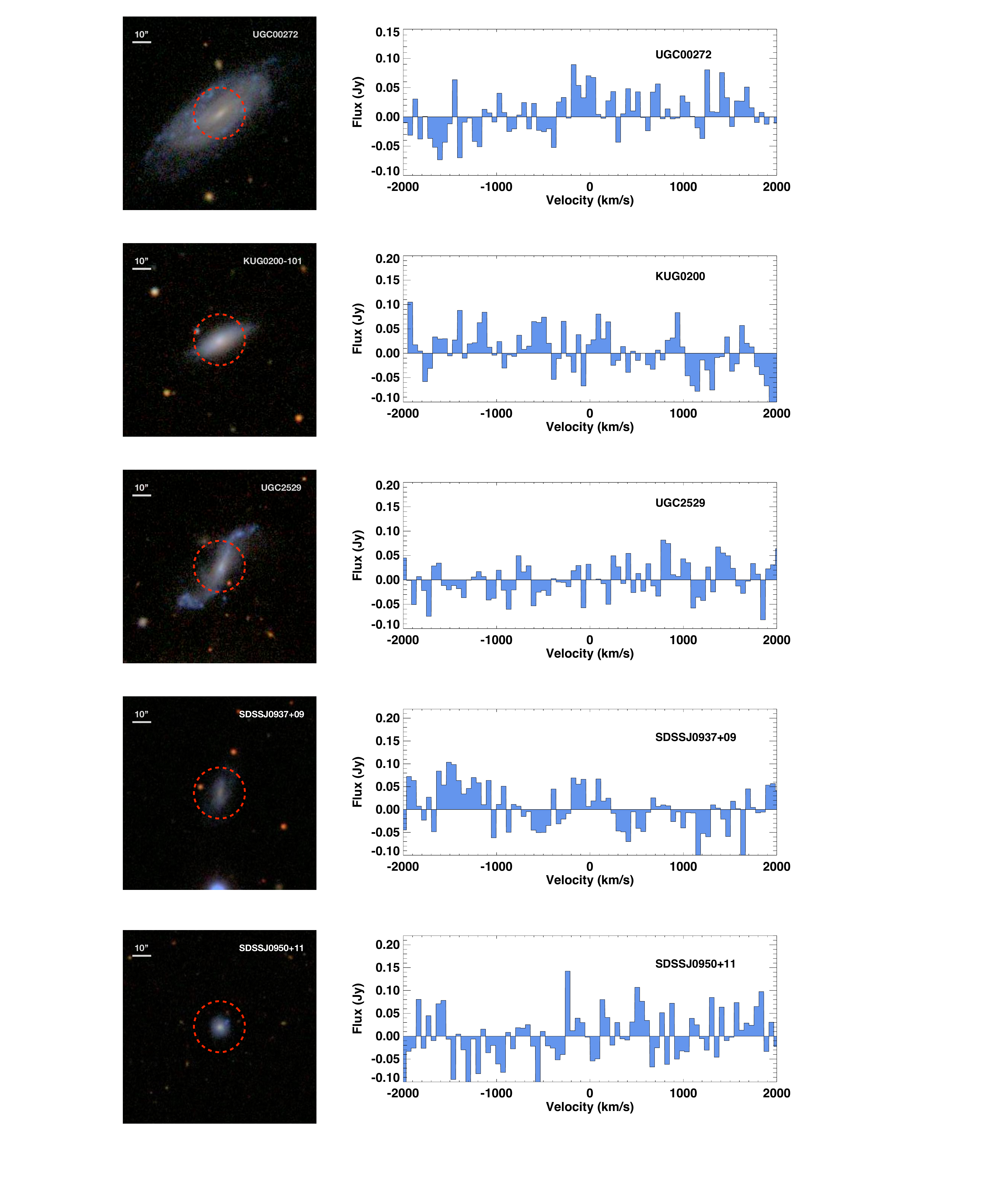}
}
\caption{SDSS cutout images of galaxies in the ALLSMOG sample {\bf not} detected in CO($2-1$) emission, and their corresponding spectra. Galaxies are shown at a constant size scale ($10''$ is shown inset in the upper left). In each left hand panel, the central circle shows the size of the $27''$ APEX beam. Spectra are presented at 60 km\,s$^{-1}$ resolution.}
\label{fig:non_mos1}
\end{figure*}

\begin{figure*}
\centering
\mbox
{
  \includegraphics[width=13cm, clip=true, trim=5cm 3cm 10cm 1cm]{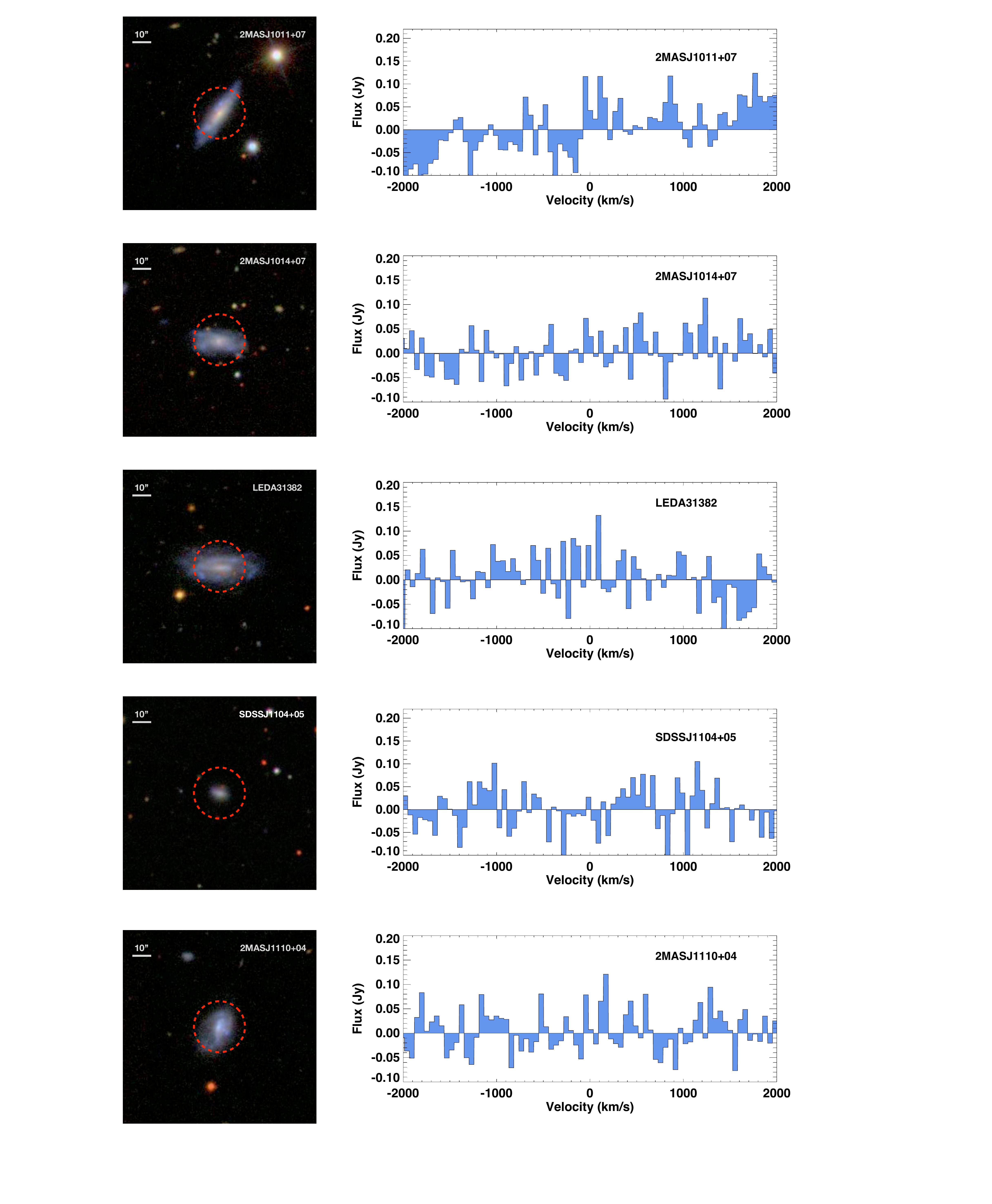}
}
\caption{Panels are the same as Fig. \ref{fig:non_mos1}. }
\label{fig:non_mos2}
\end{figure*}

\begin{figure*}
\centering
\mbox
{
  \includegraphics[width=13cm, clip=true, trim=5cm 3cm 10cm 1cm]{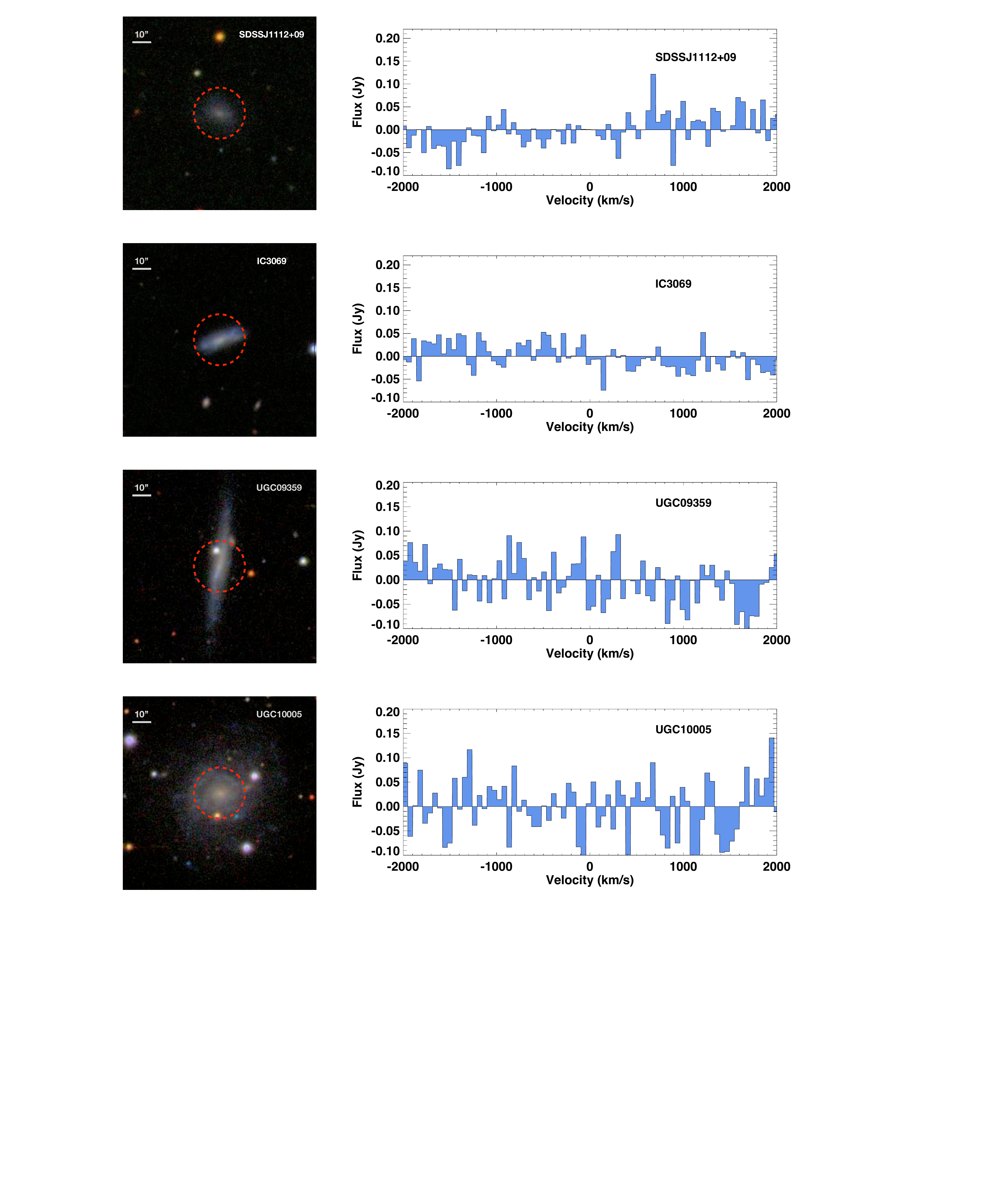}
}
\caption{Panels are the same as Fig. \ref{fig:non_mos1}. }
\label{fig:non_mos2}
\end{figure*}

\end{document}

%% file: tab1.tex
\begin{table*}
\centering
\caption{Physical properties of the ALLSMOG sample. Column (1): Source name from NED. Columns (2) and (3): RA and DEC of SDSS source. Column (4): SDSS spectral redshift. Column (5): Morphological type from NED (blank if unavailable). Column (6): Log of the B-band optical diameter, in units of 0.1 arcmin.}
  \begin{tabular}{@{}lcccccc@{}}
\hline
 (1) & (2) & (3) & (4) & (5) & (6)\\
Full name & RA & DEC & Redshift  & Morphological Type & Size   \\
          & [J2000] & [J2000] & [z] & [NED] & log(0.1)d \\
 \hline
 \hline

UGC00272                 & 00 27 49.7  &  $-$01 11 60  & 0.0130  & Scd     & 1.128 \\
IC0159                   & 01 46 25.0  &  $-$08 38 12  & 0.0130  & Scd     & 1.131	 \\
KUG0200-101              & 02 03 16.6  &  $-$09 53 26  & 0.0129  & Sbc     & 0.810 \\
UGC2004                  & 02 31 59.6  &  +00 54 36    & 0.0218  & SAB(s)c & 0.850	 \\
UGC2529                  & 03 05 29.6  &  $-$00 22 54  & 0.0248  & S       & 0.928 \\
NGC1234                  & 03 09 39.1  &  $-$07 50 46  & 0.0124  & Scd     & 1.198  \\
SDSSJ0805+06             & 08 05 37.7  &  +06 59 35	   & 0.0152  & ---     & 0.557 \\
2MASXJ08393919+0349424   & 08 39 39.2  &  +03 49 43    & 0.0266  & S0/a    & 0.642  \\
2MASJ0846+02             & 08 46 54.0  &  +02 30 05    & 0.0282  & Sa      & 0.588 \\
2MASXJ08555592+0345300   & 08 55 56.0  &  +03 45 30    & 0.0275  & ---     & 0.729 \\
2MASXJ08580528+0345228   & 08 58 05.3  &  +03 45 23    & 0.0269  & S0/a    & 0.578         \\
2MASXJ09105875+0752191   & 09 10 58.8  &  +07 52 19    & 0.0284  & Sbc     & 0.754	 \\
UGC4977                  & 09 21 59.6  &  +03 22 43    & 0.0278  & Sc      & 0.924	 \\
MCG+00-25-005            & 09 36 35.4  &  +01 07 00    & 0.0164  & Sb      & 0.855	 \\
SDSSJ0937+09             & 09 37 09.0  &  +09 27 51    & 0.0224  & ---     & 0.848  \\
NGC2936                  & 09 37 45.0  &  +02 45 34    & 0.0240  & Irr     & 0.021\\
2MASXJ09393527+0624513   & 09 39 35.3  &  +06 24 52    & 0.0249  & Scd     & 0.625	 \\
SDSSJ0950+11             & 09 50 11.2  &  +11 18 30    & 0.0188  & ---     & 0.465 \\
2MASXJ10110944+0746489   & 10 11 09.4  &  +07 46 49    & 0.0262  & Sc      & 0.843 \\
2MASXJ10145891+0748028   & 10 14 58.9  &  +07 48 03    & 0.0283  & Scd     & 0.743 \\
IC0605                   & 10 22 24.1  &  +01 11 54    & 0.0215  & Sc      & 0.887	 \\
UGC5648                  & 10 26 08.8  &  +04 22 22    & 0.0228  & S       & 1.073	 \\
LEDA31382                & 10 35 42.3  &  +05 36 58    & 0.0272  & Scd     & 0.871 \\
NGC3436                  & 10 52 28.9  &  +07 54 15    & 0.0205  & Sab     & 0.935	 \\
SDSSJ1104+05             & 11 04 14.6  &  +05 07 37	   & 0.0176  & ---     & 0.386 \\
2MASXJ11103746+0411286   & 11 10 37.4  &  +04 11 29    & 0.0289  & Sm      & 0.617 \\
SDSSJ1112+09             & 11 12 50.2  &  +09 31 39	   & 0.0209  & ---     & 0.768 \\
MCG+00-29-013            & 11 18 49.6  &  +00 37 10    & 0.0254  & Sc      & 0.761	 \\
UGC6329                  & 11 18 56.2  &  +00 10 34    & 0.0249  & Scd     & 0.905	 \\
UGC06838                 & 11 51 56.1  &  $-$02 38 33  & 0.0129  & SAab    & 0.850 \\
IC3069                   & 12 16 19.9  &  +10 09 39    & 0.0195  & ---     & 0.703 \\
SDSSJ1328$-$02           & 13 28 46.9  &  $-$02 02 28  & 0.0124  & ---     & 0.750 \\
MCG00-34-038             & 13 29 50.4  &  $-$01 25 45  & 0.0213  & Sbc     & 0.826   \\
CGCG017		            & 13 32 50.2  &  $-$03 04 58  & 0.0142  & Sbc     & 1.159\\
UGC085276                & 13 32 55.1  &  $-$01 09 34  & 0.0127  & S?      & 0.933\\
NGC5405                  & 14 01 09.5  &  +07 42 08    & 0.0230  & S?      & 0.917	 \\
NGC5414                  & 14 02 03.5  &  +09 55 46    & 0.0141  & S?      & 0.988	 \\
UGC09359                 & 14 33 15.8  &  $-$01 08 24  & 0.0138  & Sdm     & 0.926 \\
CGCG050$-$042            & 15 36 46.6  &  +07 50 01    & 0.0111  & ---     & 0.645 \\
UGC10005                 & 15 45 14.4  &  +00 46 20    &	 0.0128  & Sc      & 0.892 \\
UGC11631                 & 20 47 59.9  &  $-$00 10 48  & 0.0141  & Scd     & 0.998	 \\ 
2MASXJ22350698$-$0845550 & 22 35 07.0  &  $-$08 45 55  & 0.0238  & ---     & 0.758	 \\

\hline
\end{tabular}
\label{tab:props}
\end{table*}

%% file: tab2.tex
\begin{table*}
\centering
\caption{Details of the APEX observations. Column (1): Source name from NED. Columns (2) and (3): RA and DEC of SDSS source. Column (4): spectral rms (per 20 km/s channel). Column (5): Total time on-source, after discarding distorted scans. Column (6): Flux density, and associated error, of the $^{12}$CO$(2-1)$ line. For non-detected galaxies, $3\sigma$ upper limits are quoted. Column (7): FWHM of the CO$(2-1)$ line, obtained by fitting a Gaussian profile. (8): Percentage of the CO flux recovered by the beam (see Appendix A) }
  \begin{tabular}{@{}lccccccc@{}}
\hline
(1) & (2) & (3) & (4) & (5) & (6) & (7)& (8) \\
Name & RA & DEC  & rms$_{(20)}$ & Time on source &  S$_{CO(2-1)}$ & FHWM & Beam coverage  \\
          & [J2000] & [J2000] & [mJy] & [Hours]   & [Jy km/s]  & [km/s] & [Fraction]  \\
 \hline
 \hline
 {\it Detected sources:} \\

IC0159                  & 01 46 25.0  &   $-$08 38 12  &  69   & 1.2  & $31.1 \pm 5.8$ &  157  &  $  0.77   \pm{0.07} $ \\
UGC2004                 & 02 31 59.6  &   +00 54 36    &  131  & 0.8  & $114.9\pm12.5$ &  212  &  $  0.93   \pm{0.03} $ \\
NGC1234                 & 03 09 39.1  &   $-$07 50 46  &  45   & 2.9  & $14.0 \pm 3.5$ &  132  &  $  0.71   \pm{0.08} $ \\
2MASJ0839+03            & 08 39 39.2  &   +03 49 43    &  96   & 0.6  & $30.0 \pm 7.0$ &  18   &  $  0.99   \pm{0.01} $ \\
2MASJ0846+02            & 08 46 54.0  &   +02 30 05	   &  103  & 0.7  & $31.9 \pm 5.4$ &  36   &  $  0.99   \pm{0.01} $ \\
2MASJ0855+03            & 08 55 56.0  &   +03 45 30    &  41   & 0.8  & $29.7 \pm 17.6$&  120  &  $  0.98   \pm{0.02} $ \\
2MASJ0858+03            & 08 58 05.3  &   +03 45 23    &  45   & 1.4  & $16.5 \pm 5.9$ &  41   &  $  0.99   \pm{0.01} $ \\
2MASJ0910+07            & 09 10 58.8  &   +07 52 19    &  90   & 0.5  & $40.3 \pm 8.3$ &  188  &  $  0.97   \pm{0.02} $ \\
UGC4977                 & 09 21 59.6  &   +03 22 43    &  52   &1.6   & $32.2 \pm 5.2$ &  216  &  $  0.92   \pm{0.04} $ \\
MCG00-25-005            & 09 36 35.4  &   +01 07 00    &  110  & 0.4  & $43.6 \pm 8.7$ &  139  &  $  0.93   \pm{0.03} $ \\
NGC2936                 & 09 37 45.0  &   +02 45 34	   &  160  & 0.3  & $416.4\pm10.8$ &  164  &  $  1.00   \pm{0.00} $ \\
2MASJ0939+06            & 09 39 35.3  &   +06 24 52    &  53   & 1.2  & $30.4 \pm 5.6$ &  247  &  $  0.99   \pm{0.01} $ \\
IC0605                  & 10 22 24.1  &   +01 11 54    &  76   & 1.0  & $37.5 \pm 6.8$ &  174  &  $  0.90   \pm{0.04} $ \\
UGC5648                 & 10 26 08.8  &   +04 22 22    &  69   & 0.9  & $28.4 \pm 6.5$ &  195  &  $  0.84   \pm{0.07} $ \\
NGC3436                 & 10 52 28.9  &   +07 54 15    &  78   & 0.5  & $23.8 \pm 5.2$ &  98   &  $  0.88   \pm{0.05} $ \\
MCG00-29-013            & 11 18 49.6  &   +00 37 10    &  82   & 0.9  & $43.1 \pm 7.6$ &  188  &  $  0.94   \pm{0.03} $ \\
UGC6329                 & 11 18 56.2  &   +00 10 34    &  47   & 1.6  & $11.1 \pm 3.0$ &  91   &  $  0.87   \pm{0.05} $ \\
UGC06838                & 11 51 56.1  &   $-$02 38 33  &  58   & 1.5  & $7.8  \pm 3.0$ &  51   &  $  0.95   \pm{0.03} $ \\
MCG00-34-038	          & 13 29 50.4  &   $-$01 25 45  &  156  & 0.3  & $84.5 \pm 8.5$ &  84   &  $  0.94   \pm{0.03} $ \\
CGCG017		          & 13 32 50.2  &   $-$03 04 58  &  100  & 0.8  & $49.3 \pm 9.7$ &  90   &  $  0.72   \pm{0.07} $ \\
UGC085276               & 13 32 55.1  &   $-$01 09 34  &  101  & 0.8  & $41.1 \pm 7.0$ &  58   &  $  0.87   \pm{0.04} $ \\
NGC5405                 & 14 01 09.5  &   +07 42 08    &  109  & 0.5  & $20.2 \pm 4.6$ &  37   &  $  0.83   \pm{0.07} $ \\
NGC5414                 & 14 02 03.5  &   +09 55 46    &  83   & 0.8  & $49.3 \pm 9.3$ &  91   &  $  0.86   \pm{0.05} $ \\
UGC11631                & 20 47 59.9  &   $-$00 10 48  &  35   & 4.4  & $29.5 \pm 2.9$ &  157  &  $  0.89   \pm{0.05} $ \\
2MASJ2235$-08$          & 22 35 07.0  &   $-$08 45 55  &  49   & 3.0  & $27.1 \pm 3.4$ &  106  &  $  0.96   \pm{0.02} $ \\

\\
 {\it Non-detected sources:} \\
 
UGC00272                &  00 27 49.7  &  $-$01 11 60 & 63  & 0.9  & $<9.6$ & --- 	&  $  0.79   \pm{0.07} $ \\
KUG0200                 &  02 03 16.6  &  $-$09 53 26 & 62  & 1.1  & $<9.5$ & --- 	&  $  0.96   \pm{0.03} $ \\
UGC2529                 &  03 05 29.6  &  $-$00 22 54 & 64  & 1.2  & $<9.8$ & --- 	&  $  0.93   \pm{0.04} $ \\
SDSSJ0805+06            &  08 05 37.7  &  +06 59 35	  & 123 & 0.8  & $<18.8$ & ---   &  $  0.99   \pm{0.01} $ \\
SDSSJ0937+09            &  09 37 09.0  &  +09 27 51	  & 71  & 0.8  & $<10.9$ & --- 	&  $  0.93   \pm{0.03} $ \\
SDSSJ0950+11            &  09 50 11.2  &  +11 18 30	  & 91  & 0.8  & $<13.9$ & --- 	&  $  0.99   \pm{0.01} $ \\
2MASJ1011+07            &  10 11 09.4  &  +07 46 49   & 63  & 0.8  & $<9.6$ & --- 	&  $  0.95   \pm{0.03} $ \\
2MASJ1014+07            &  10 14 58.9  &  +07 48 03   & 80  & 0.4  & $<12.2$ & ---   &  $  0.97   \pm{0.01} $ \\
LEDA31382               &  10 35 42.3  &  +05 36 58   & 76  & 0.5  & $<11.6$ & --- 	&  $  0.93   \pm{0.03} $ \\
SDSSJ1104+05            &  11 04 14.6  &  +05 07 37	  & 91  & 1.0  & $<13.9$ & --- 	&  $  0.99   \pm{0.01} $ \\
2MASJ1110+04            &  11 10 37.4  &  +04 11 29   & 67  & 0.5  & $<10.3$ & ---   &  $  0.99   \pm{0.01} $ \\
SDSSJ1112+09            &  11 12 50.2  &  +09 31 39	  & 60  & 1.5  & $<9.2$ & ---  	&  $  0.96   \pm{0.02} $ \\
IC3069                  &  12 16 19.9  &  +10 09 39	  & 52  & 1.6  & $<7.9$ & --- 	&  $  0.98   \pm{0.01} $ \\
SDSSJ1328$-$02          &  13 28 46.9  &  $-$02 02 28 & 160 & 0.5  & $<24.5$ & --- 	&  $  0.97   \pm{0.02} $ \\
UGC09359                &  14 33 15.8  &  $-$01 08 24 & 76  & 1.0  & $<11.6$ & --- 	&  $  0.93   \pm{0.05} $ \\
CGCG050-042             &  15 36 46.6  &  +07 50 01   & 270 & 0.3  & $<41.3$ & --- 	&  $  0.98   \pm{0.02} $ \\
UGC10005                &  15 45 14.4  &  +00 46 20   & 90  & 1.4  & $<13.8$ & --- 	&  $  0.89   \pm{0.05} $ \\
\hline
\end{tabular}
\label{tab:obs}
\end{table*}

%% file: tab3.tex
\begin{table*}
\centering
\caption{Derived parameters of the ALLSMOG sample. Column (1): Source name from NED. Column (2): Luminosity of the CO($2-1$) emission line. Column (3): Molecular hydrogen mass, calculated assuming a constant $\alpha_{\rm CO}$ appropriate for the Milky-Way.  Column (4): Molecular hydrogen mass, calculated assuming a metallicity-dependent $\alpha_{\rm CO}$, as described in the text. Column (5): Atomic hydrogen mass. Column (6): Stellar mass, calculated from SDSS. Column (7): Gas-phase metallicity, calculated as described in the text. Column (8): Log star formation rate, from SDSS. }
  \begin{tabular}{@{}lccccccc@{}}
\hline
(1) & (2) & (3) & (4) & (5) & (6) & (7) & (8) \\
Name &  L$'_{CO(2-1)}$    & log M(H$_2)^{\rm \alpha = const}$ & log M(H$_2)^{\rm \alpha= f(Z)}$ & log M(\HI) & log M* & Metallicity & log SFR\\
          & [$10^8$ K km/s pc$^2$] & [M$_{\sun}$] & [M$_{\sun}$] & [M$_{\sun}$] & [M$_{\sun}$]  & [12+log(O/H)] & [M$_{\sun}$/yr]\\
 \hline
 \hline
 {\it Detected sources:} \\
 
IC0159                              & $0.76\pm0.12$ & $8.53 \pm 0.07$ & $8.38 \pm 0.07$ & 9.55 & 9.70  & 8.89 & 0.23 \\
UGC2004                             & $6.43\pm0.67$ & $9.46 \pm 0.04$ & $9.23 \pm 0.04$ & 9.83 & 9.61  & 9.15 & 0.36 \\
NGC1234                             & $0.33\pm0.07$ & $8.18 \pm 0.09$ & $7.99 \pm 0.09$ & 9.67 & 9.58  & 8.98 & 0.21 \\
2MASJ0839+03                        & $2.33\pm0.54$ & $9.02 \pm 0.09$ & $8.81 \pm 0.09$ & ---  & 9.94  & 9.08 & 0.58 \\
2MASJ0846+02                        & $2.77\pm0.47$ & $9.09 \pm 0.07$ & $8.90 \pm 0.07$ & ---  & 9.84  & 8.99 & 0.09 \\
2MASJ0855+03                        & $2.49\pm1.47$ & $9.05 \pm 0.20$ & $8.94 \pm 0.20$ & ---  & 9.73  & 8.81 & 0.06\\
2MASJ0858+03                        & $1.30\pm0.46$ & $8.76 \pm 0.13$ & $8.55 \pm 0.13$ & ---  & 9.80  & 9.08 & 0.11\\
2MASJ0910+07                        & $3.64\pm0.73$ & $9.21 \pm 0.08$ & $8.99 \pm 0.08$ & 9.43 & 9.85  & 9.11 & 0.07 \\
UGC4977                             & $2.94\pm0.47$ & $9.12 \pm 0.07$ & $8.94 \pm 0.07$ & ---   & 9.91  & 8.95 & 0.19 \\
MCG00-25-005                        & $1.39\pm0.26$ & $8.79 \pm 0.07$ & $8.60 \pm 0.07$ & 10.1 & 9.70  & 9.01 & 0.02 \\
NGC2936                             & $26.0\pm0.69$ & $10.0 \pm 0.01$ & $9.83 \pm 0.01$ & ---  & 8.53  & 9.17 & 0.34 \\
2MASJ0939+06                        & $2.07\pm0.37$ & $8.96 \pm 0.07$ & $8.92 \pm 0.07$ & 9.81 & 9.44  & 8.71 & 0.33 \\
IC0605                              & $2.12\pm0.35$ & $8.98 \pm 0.07$ & $8.79 \pm 0.07$ & 9.93 & 9.89  & 8.97 & 0.23 \\
UGC5648                             & $1.92\pm0.42$ & $8.93 \pm 0.08$ & $8.72 \pm 0.08$ & 10.1 & 9.99  & 9.08 & 0.02 \\
NGC3436                             & $1.25\pm0.25$ & $8.75 \pm 0.08$ & $8.60 \pm 0.08$ & 9.77 & 9.64  & 8.89 & 0.05 \\
MCG00-29-013                        & $3.21\pm0.54$ & $9.16 \pm 0.07$ & $8.95 \pm 0.07$ & 9.66 & 9.98  & 9.07 & 0.15 \\
UGC6329                             & $0.86\pm0.21$ & $8.58 \pm 0.10$ & $8.40 \pm 0.10$ & 10.0 & 9.80  & 8.98 & 0.02 \\
UGC06838                            & $0.13\pm0.05$ & $7.78 \pm 0.15$ & $7.62 \pm 0.15$ & 9.59 & 9.32  & 8.91 & 0.33 \\
MCG00-34-038                        & $4.44\pm0.43$ & $9.30 \pm 0.04$ & $9.08 \pm 0.04$ & 9.96 & 9.91  & 9.09 & 0.14 \\
CGCG017                             & $1.53\pm0.27$ & $8.83 \pm 0.08$ & $8.61 \pm 0.08$ & 10.0 & 9.96  & 9.12 & $-0.79$ \\
UGC085276                           & $0.84\pm0.12$ & $8.57 \pm 0.07$ & $8.38 \pm 0.07$ & 9.31 & 9.68  & 9.01 & $-0.44$ \\
NGC5405                             & $1.40\pm0.27$ & $8.80 \pm 0.08$ & $8.60 \pm 0.08$ & 10.1 & 9.98  & 9.03 & 0.27 \\
NGC5414                             & $1.24\pm0.22$ & $8.74 \pm 0.07$ & $8.58 \pm 0.07$ & 9.79 & 9.74  & 8.92 & 0.25 \\
UGC11631                            & $0.72\pm0.06$ & $8.51 \pm 0.04$ & $8.41 \pm 0.04$ & 9.72 & 9.57  & 8.80 & 0.59 \\ 
2MASJ2235$-08$                      & $1.74\pm0.21$ & $8.89 \pm 0.05$ & $8.70 \pm 0.05$ & 9.75 & 9.78  & 9.00 & 0.25 \\
\\
 {\it Non-detected sources:} \\
UGC00272       & $<0.22$ & $<7.85$ & $<7.81$ & 9.69 & 9.36 & 8.91 & 0.21\\
KUG0200        & $<0.18$ & $<7.96$ & $<7.93$ & 9.63 & 9.11 & 8.61 & 0.92\\
UGC2529        & $<0.70$ & $<8.43$ & $<8.44$ & 10.1 & 9.42 & 8.74 & 0.02\\
SDSSJ0805+06   & $<0.48$ & $<8.33$ & $<8.32$ & 9.18 & 8.62 & 8.66 & $-1.25$ \\
SDSSJ0937+09   & $<0.64$ & $<8.47$ & $<8.44$ & 9.77 & 8.75 & 8.64 & $-0.82$ \\
SDSSJ0950+11   & $<0.54$ & $<8.43$ & $<8.41$ & 8.80 & 8.57 & 8.61 & $-0.88$ \\
2MASJ1011+07   & $<0.75$ & $<8.40$ & $<8.38$ & 9.83 & 9.96 & 8.85 & 0.11\\
2MASJ1014+07   & $<1.09$ & $<8.56$ & $<8.54$ & 9.74 & 9.51 & 8.84 & 0.00\\
LEDA31382      & $<1.00$ & $<8.52$ & $<8.50$ & 9.95 & 9.95 & 8.85 & 0.08\\
SDSSJ1104+05   & $<0.47$ & $<8.46$ & $<8.41$ & 8.94 & 8.67 & 8.52 & $-1.04$ \\
2MASJ1110+04   & $<0.94$ & $<8.52$ & $<8.51$ & 9.73 & 9.51 & 8.80 & 0.56\\
SDSSJ1112+09   & $<0.45$ & $<8.36$ & $<8.32$ & 9.35 & 8.89 & 8.60 & $-0.85$ \\
IC3069         & $<0.33$ & $<8.23$ & $<8.20$ & 9.32 & 8.99 & 8.60 & $-0.51$ \\
SDSSJ1328$-$02 & $<0.43$ & $<8.09$ & $<8.02$ & 9.63 & 8.82 & 9.00 & $-0.50$ \\
UGC09359       & $<0.26$ & $<8.07$ & $<8.07$ & 9.58 & 8.96 & 8.65 & 0.17\\
CGCG050-042    & $<0.57$ & $<8.51$ & $<8.47$ & 9.03 & 8.54 & 8.55 & $-1.44$ \\
UGC10005       & $<0.28$ & $<8.00$ & $<7.97$ & 9.45 & 9.06 & 8.79 & 1.08\\
\hline
\end{tabular}
\label{tab:phys}
\end{table*}

%{\it  \hspace{2pt} 2MASJ0846+02\_a} & $2.69\pm0.50$ & $8.97 \pm 0.07$ & & $''$ & $''$  & $''$ & $''$ \\
%{\it  \hspace{2pt} 2MASJ0846+02\_b} & $0.98\pm0.62$ & $8.53 \pm 0.21$ & & $''$ & $''$  & $''$ & $''$ \\
%{\it \hspace{2pt} NGC2936\_a}       & $22.9\pm0.95$ & $9.90 \pm 0.02$ & & $''$ & $''$  & $''$ & $''$ \\
%{\it \hspace{2pt} NGC2936\_b}       & $5.90\pm0.44$ & $9.31 \pm 0.03$ & & $''$ & $''$  & $''$ & $''$ \\
%{\it  \hspace{2pt} MCG00-29-013\_a} & $1.65\pm0.36$ & $8.76 \pm 0.05$ & & $''$ & $''$  & $''$ & $''$ \\
%{\it  \hspace{2pt} MCG00-29-013\_b} & $1.22\pm0.25$ & $8.63 \pm 0.08$ & & $''$ & $''$  & $''$ & $''$ \\